\documentclass[12pt, preprint]{aastex}
\usepackage{lscape}

\slugcomment{To appear in the Astronomical Journal}

\shorttitle{Sizes of the Nearest Young Stars}
\shortauthors{McCarthy and White}

\begin{document}


\title{The Sizes of the Nearest Young Stars}


\author{Kyle McCarthy$^{1,2}$ and Russel J. White$^{1}$}
\affil{$^{1}$Department of Physics and Astronomy, Georgia State 
University, Atlanta, GA 30303-4106
       $^{2}$Department of Physics and Astronomy, University of 
Kentucky, Lexington, KY 40508-0055}

\email{kyle.mccarthy@uky.edu}




\begin{abstract}
We present moderate resolution (R $\sim$ 3575) optical spectra of 19
known or suspected members of the AB Doradus and $\beta$ Pictoris 
Moving Groups, obtained with the DeVeny Spectrograph on the 72-inch Perkins
telescope at Lowell Observatory.  For 4 of 5 recently proposed members, 
signatures of youth such as Li\,I 6708 \AA\, absorption and H$\alpha$ 
emission further strengthen the case for youth and membership.  The lack
of detected lithium in the proposed $\beta$ Pic member TYC 2211-1309-1
implies that it is older than all other K-type members, and weakens the
case for membership.  Effective temperatures are determined via line 
ratio analyses for the 11 F, G and early K stars observed, and via spectral 
comparisons for the 8 late-K and M stars observed.  We assemble updated 
candidate membership lists for these Moving Groups that account for known 
binarity.  Currently the AB Dor moving group contains 127 proposed members 
and the $/beta$ Pic moving group holds 77 proposed members.  We then use 
temperature, luminosity, and distance estimates to 
predict angular diameters for these stars; the motivation is to identify stars that can be 
spatially resolved with long-baseline optical/infrared interferometers in 
order to improve age estimates for these Groups and to constrain 
evolutionary models at young ages. Considering the portion of the sky 
accessible to northern hemisphere facilities (DEC $> -30$), 6 stars have 
diameters large enough to be spatially resolved ($\theta > 0.4$ mas) with the 
CHARA Array, which currently has the world's longest baseline of 331-m; 
this subsample includes the low mass M2.5 member of AB Dor, GJ 393, which 
is likely to still be pre-main sequence.  For southern hemisphere facilities 
(DEC $< +30$), 18 stars have diameters larger than this limiting size, 
including the low 
mass debris disk star AU Mic (0.72 mas).  However, the longest baselines 
of southern hemisphere interferometers (160-m) are only able to resolve 
the largest of these, the B6 star $\alpha$ Gru (1.17 mas); proposed 
long-baseline stations may alleviate the current limitations.
\end{abstract}



\section{Introduction}
Stellar evolutionary models are an essential tool in the study of star
and planet formation since, in most cases, they provide the only means 
of estimating fundamental stellar properties like mass and age. Despite 
decades of progress with these models, there remain considerable 
uncertainties in the requisite input physics \citep[]{bar2002,you2005}. 
For main-sequence stars, input parameters such as mixing length or 
abundances can be tuned to predict 
properties consistent with observations;
this is in large part because of the wealth of precise dynamical masses
and stellar radii that high resolution techniques such as interferometry
have provided \citep[e.g.][]{mic1921,han1974,lab1975}. Unfortunately,
evolutionary models are considerably more uncertain at pre-main sequence
ages. While this is partially due to the additional complicating
properties of young stars like rapid rotation and star spots, it is 
also a consequence of most star forming regions being too distant ($>$
100 pc) to permit detailed interferometric measurements of their
members. As a consequence, mass estimates for young stars vary by as much as
50\%-200\% and age estimates vary by up to a factor of 10, depending
upon the adopted evolutionary model \citep[e.g.][]{white2004, mat2007}.

Fortunately, two advances in observational astronomy over the last decade
are enabling new fundamental measurements of young stars.  The first is
the discovery of many young ($\lesssim 100$ Myr) stars within close 
proximity of the Sun \citep[e.g.][]{zuc2004a,lop2006,tor2008,lep2009}.
The majority of these young stars are members of small, slowly dispersing
T Associations, now commonly referred to as Moving Groups.  As cataloged 
in the recent review article by \citet{tor2008}, there are now 9 well 
defined Moving Groups, all of which have ages less than $\sim$ 100 Myr 
and 4 of which have central distances within 60 pc of the Sun. Because of 
their close proximity, many of these stars also have well determined
distances from the Hipparcos parallax mission \citep{per2009}, which enable 
much more precise stellar luminosity and size estimates.

An important complement to these discoveries are rapid advances in
long-baseline interferometry.  Facilities now operate at infrared and even
optical wavelengths and can achieve spatial resolutions better than 1
milliarcsecond (mas).  For comparison, at a distance of 10 pc, the sun would 
have and angular diameter of 0.93 mas.  Currently, the longest baseline optical/infrared
interferometer is Georgia State University's CHARA Array, located on
Mt. Wilson in California \citep{ten2005}. With a long baseline of 331-m, 
the Array yields a minimum angular resolution below 1 mas in the infrared 
\citep[e.g.][]{boy2008,bai2010}; the smallest published angular diameter 
measured by CHARA is that of HD 189733 with an angular size of 
0$\farcs$377 $\pm$ 0.024 mas in the H-band \citep{bai2007}.  
This resolution limit will continue to improve as instruments that operate 
optical wavelengths are utilized \citep[e.g.][]{mou2010}.  Other 
operational long-baseline
interferometers include the Navy Prototype Optical Interferometer in
Arizona \citep{hum2003}, the Very Large Telescope Interferometer
\citep{pet2007} located on Cerro Paranal, Chile, and the Sydney
University Stellar Interferometer \citep{dav1999} in New South Wales,
Australia.  Although these interferometers have thus far utilized baselines 
less than 160-m, the proposed extension to longer baselines
and observations at optical wavelengths suggest they will achieve
similar spatial resolutions to the CHARA Array in the near future.

The primary purpose of this article is to investigate which, if any,
of the nearest young stars are large enough and bright enough to be
spatially resolved by the current generation of optical/infrared
interferometers. We also comment on the types of improvements that 
would dramatically increase the number of young stars that could be 
spatially resolved.

\section{The Nearest Young Stars}

In this analysis, we focus on the two nearest Moving Groups,
AB Doradus and $\beta$ Pictoris.  \citet{tor2008} list mean distances
of 34 and 31 pc with dispersions of 26 and 31 pc for these groups, 
respectively. These distances are $\sim 30\%$ closer than the next closest 
Moving Groups, Tucana/Horologium and TW Hydrae, both at an average 
distance of 48 pc.  In addition, unlike the Tucana/Horologium and 
TW Hydrae Moving Groups, both AB Doradus and $\beta$ Pictoris have 
many members in the northern hemisphere and thus are accessible to the 
CHARA Array.

\emph{AB Doradus (AB Dor)}: 
Age estimates for the AB Dor Moving Group range from being comparable
to the youngest known open clusters \citep[30-50 Myr;][]{zuc2004b, 
clo2005} to 
being coeval with the Pleiades \citep[100-125 Myr;][]{ort2007,luh2005}.
Based on an analysis of lithium abundances of known members, 
\cite{sil2009} favor an intermediate age of 70 Myr (see also Janson
et al. 2007).

In Table 1 we provide a current list of 127 potential members of this 
Moving Group, along with distance estimates, Johnson $V$-band magnitudes, 
spectral types, $K_{2MASS}$ magnitudes, assigned temperatures and 
multiplicity status.  
The majority of these members come from the compilation paper of 
\citet{tor2008}, who create a membership list based upon the previous 
studies of \citet{cov1997,zuc2004b,zic2005,lop2006}.  
\citet{tor2008}
cite a membership probability for each star based on evolutionary and
kinematic criteria, as described in \citet{tor2006}.  The membership
probabilities span from 60\% to 100\%.  
To this list we added the 6 proposed members from \citet{sch2010},
who identify members based on trigonometric distances, X-ray 
emission, H$\alpha$ emission, and radial velocity measurements.  Finally, 
we add 3 candidate members from \citet{sil2009}\footnote{The newly
proposed AB Dor member BD+09 412 by \citet{sil2009} is also HW Cet; the 
SIMBAD database does not currently recognize these 2 stars as being the 
same star.} who use the 
same criteria as \citet{tor2008} to identify candidate members, and  7 candidates 
from \citet{zuc2011} who use galactic space motions, locations on a 
color-magnitude diagram, lithium abundances, and X-ray luminosities to 
identify candidate members; the 7 stars identified in \citet{zuc2011} 
are all A- or late B-type stars.

Of these 127 stars, 54 have distances determined from Hipparcos 
parallax measurements \citep[][marked with an H in Table 1]{lee2007}.  
Another 40 stars have distances estimated kinematically by \citet{tor2008} 
or \citet{sch2010}.  Any companions (defined below) to these stars are 
assumed to be at the same distance.  Only 3 of the 127 stars have
no distance estimates (HW Cet, TYC 0091-0082-1, RX J0928.5-7815).  For 
these stars, we estimate their distances by assuming they have the same 
radii (calculated below) as stars of the same spectral type, and then 
determine the distances that would yield a luminosity consistent with
their observed $V$ magnitudes; these calculated distances are marked in 
the Tables.

The Johnson $V$-band magnitudes are calculated in the majority of cases (97
stars) from $V_{T}$ magnitudes in the Tycho-2 Catalog \citep{hog2000},
using the prescription of \citet{bes2000}, which relies on knowing the
$B_{T}$ and $V_{T}$ magnitudes.  For the 4 stars that only have a 
$V_{T}$ magnitude we estimate their ($B_{T}$ - $V_{T}$) color using their 
spectral type and the color relations of \citet{har1994}.  In the absence 
of any Tycho-2 magnitudes, $V$-band magnitudes are assembled from 
measurements in the literature, when available.  Stars with literature
$V$ magnitudes include AB Dor BaBb \citep{cam1997}, GSC 8894-0426 
\citep{cra1997}, GSC 8544-1037, CD-45 14955 \citep{tor2006, tor2008}, 
and $\alpha$ Gru and $\delta$ Scl \citep{zuc2011}.  In total 107 of
the 127 potential members have $V$-band measurements.  We note that 
the $V$ magnitudes of 14 stars represent the combined light of 2 or more 
stars (values listed in brackets in the Tables).

Of the assembled 2MASS\footnote{This publication 
makes use of data products from the Two Micron All Sky Survey, which is a 
joint project of the University of Massachusetts and the Infrared Processing 
and Analysis Center/California Institute of Technology, funded by the 
NASA and the NSF.}
$K$-band magnitudes, 17 are of 2 or more stars (value listed
in brackets).  Fewer pairs have resolved 2MASS measurements than $V$-band
measurements because of the higher resolution of the Tycho instrument
\citep[$\sim$ 0\farcs8;][]{hog2000} compared to the 2MASS survey \citep[$\sim$ 
3\farcs;][]{skr2006}.  Although the candidate brown dwarf 
companion to CD-35 2722 A \citep{wah2011} is spatially unresolved at
$V$ and $K$, we do not mark the measurement as ``combined'', given the very low
flux ratio measured for this pair at infrared wavelengths.

The assembled spectral types are as listed in \citet{tor2008} or the 
discovery paper for more recent additions.  These spectral types are 
used to estimate temperatures following the temperature-spectral type 
scale in \citet{har1994} for F, G, K, and M stars and 
\citet{ken1995} for B and A stars.  There are 4 stars without a 
spectral type (BD+23 296 B, HW Cet, AK Pic B, and CD-49 2843 B), but 
fortunately all have Tycho-2 $B_{T}$ and $V_{T}$ 
from which spectral types and temperatures can be estimated using the
above color relations.  These estimated spectral types are marked with
a colon in Tables 1 and 2.

In the last column of Table 1 we describe the multiplicity status of
each star (single, binary, triple, quadruple), if 
known\footnote{\citet{tor2008} provides a useful Table summarizing known
binaries in AB Dor, to which we note the following corrections.
UY Pic AB is listed with a separation of 10\farcs3 in their Table 14, 
but its separation is 18\farcs3 in the 2MASS database.  The 16\farcs2 companion listed to HD 
45270 can not be confirmed; we do not include this companion in our list.
We assume that the spectroscopic companion to PX Vir reported in \citet{tor2008} is
the same $\sim$ 0\farcs4 companion spatially resolved by \citet{eva2011}.}.  
Stars are only listed as single if they have been included in high
spatial resolution imaging surveys \citep{kas2007,bil2010,nie2010,eva2011} 
and no companion was identified.  For multiples, projected separations 
are given if the pair has been spatially resolved.  The separations
for pairs closer than $\sim$ 10\farcs0 are as assembled in \citet{tor2008},
while wider pairs are calculated from the stars' 2MASS positions.

We classify 26 systems as multiple within the AB Dor Moving Group by 
assuming that all systems with projected separations less than 75\farcs0 
are physically associated.  This separation limit is set 
in order to classify the $\beta$ Pic triples 51 Eri \& GJ 3305 AB and
HR 7012 AB \& CD-64 1208 B as physical systems.  Although the separation
limit is somewhat subjective, the projected separations in these systems 
correspond to distances of 1979 AU and 2163 AU, which is consistent with 
many known physically bound binaries \citep{rag2010}, while
all wider pairs have projected separations greater than 12,000 AU.
We nevertheless note that CD-60 1425 \& GSC 8894-0426 (27.2 arcmin), 
BD+08 4561 \& TYC 1090-543-1 (4.1 arcmin) and AK Pic AB \& CD-61 1439 
(13.4 arcmin) are all spatially close on the sky.

\emph{$\beta$ Pictoris ($\beta$ Pic)}: 
The age of this cluster is more accurately determined than that of 
AB Dor, primarily because this Moving Group is younger and thus more 
distinct from the zero age main sequence. Studies of its ensemble
population suggest ages that range from 10 to 21 Myr
\citep[][but, cf. Yee $\&$ Jensen 2010, L\'{e}pine $\&$ Simon 2009, 
MacDonald $\&$ Mullan  
2010]{zuc2001, fei2006, men2008, sil2009}.

In Table 2 we provide a current list of 77 potential members of this 
Moving Group along with distance estimates, Johnson $V$-band magnitudes, 
$K_{2MASS}$ magnitudes, spectral types, assigned temperatures and 
multiplicity status.  The majority of these members come from the 
compilation of \citet{tor2008}\footnote{\citet{tor2008} mislabel the
$\beta$ Pic star HD14082 as HD 14062 in their Table 3.}, who created a 
membership list based 
upon the previous studies of \citet{moo2006} and  \citet{zuc2004a}.  
More recently, L\'{e}pine $\&$ Simon et al. (2009), Schlieder et al. 
(2010) and Kiss et al. (2011) have identified another 13 candidate 
members which we include in our sample. L\'{e}pine $\&$ Simon et al. 
(2009) and Schlieder et al. (2010) used methods described above for
identifying members, while Kiss et al. (2011) relied on 3 age 
diagnostics: color-magnitude diagrams, X-ray emission and lithium 
abundance.

Of these 77 potential members, 28 stars have distances determined from 
Hipparcos parallax measurements (marked with an H in Table 2), and 27 have 
distances determined kinematically by \citet{tor2008} or \citet{lep2009} 
or Kiss et al. (2011) or \citet{sch2010}.  Companion stars are again 
assumed to have the same distance as the primary star (marked with an 
asterisk).  Johnson $V$-band magnitudes are estimated in the majority of 
cases (47 stars) from the Tycho-2 photometry, as
described above.  
In the absence 
of Tycho-2 magnitudes, $V$-band magnitudes are assembled from the literature
for several stars including
AG Tri A, 
AG Tri B, 
GJ 3305 AB,
V824 Ara C,
GSC 8350-1924 AB,
CD-51 11312 B,
GSC 7396-0759,
CD-64 1208 AB,
1 RXS J195602.8-320720,
1RXS J200136.9-331307, 
AT Mic A, 
AT Mic B, 
WW PsA, 
TX PsA
\citep{tor2006}, and V343 Nor B \citep{son2003}.  
Fifteen stars do not have $V$-band measurements.

Eleven of the $V$-band measurements and 15 of the $K_{2MASS}$ measurements
represent light from 2 or more stars, and are marked with brackets in Table 2;
photometry for the 2 stars that harbor candidate brown dwarfs, 
PZ Tel \citep{bil2010} and $\eta$ Tel \citep{low2000} are not marked because
of their very low flux ratios.
Temperatures are estimated for all stars using the assembled spectral 
types or Tycho color (for BD+17 232 B and HR 6749 B) using the same 
prescription described above.

There are 23 known multiple star systems within the $\beta$ Pic Moving
Group using the same separation limit adopted above ($<$ 75\farcs).  
Although \citet{tor2008} list BD+05 378 (HD 12545) as a spectroscopic binary, 
as reported by \citet{son2003}, more recent radial velocity observations
show that it is not \citep{bai2011}.
Although not classified as multiples, V4046 Srg AB \& GSC 7396-0759
(2.8 arcmin), $\eta$ Tel AB \& $\eta$ Tel C (6.9 arcmin), AT Mic AB \& 
AU Mic (77.9 arcmin), and HD 199143 \& AZ Cap (5.4 arcmin) are spatially
close on the sky.

\section{Observations and Data Reduction}
Constraining evolutionary models depends critically upon having
accurately determined stellar properties, such as temperature and 
luminosity, for comparisons with model predictions.  This in 
part motivated the acquisition of optical spectra of Moving Group 
members.  In particular, we obtained moderate resolution optical 
spectra of 19 stars in the $\beta$ Pic and AB Dor Moving Groups.  
These stars are listed in Table 3.  Thirteen of these stars are 
well established members \citep[probability $>$ 95 \%;][]{tor2008},
one star has a more questionable membership (HD 15115; 60\% probability),
and 5 stars are only recently identified candidate members from
\citet[][
TYC 1186-706-1, 
TYC 2211-1309-1,
TYC 7443-1102-1,
1RXS J19506.8-3320720]{lep2009} and \citet[][TYC 1752-63-1]{sch2010}. 
In addition to these known or candidate young stars, we also obtained 
spectra of many stars with well known 
spectral types and/or effective temperatures from \citet{kir1991,gra2001,val2005}. 
All observations were obtained with the DeVeny spectrograph 
on the Perkins 72-inch Telescope at the Lowell Observatory during 3 
observing runs: 2009 Feb 5, 2009 May 7, and 2009 Aug 1 - 3.  On each
night, we also obtained sets of bias images, dome-illuminated flat 
field spectra, and Neon-Argon lamp spectra for calibration purposes.

The spectrograph was used in combination with a 1200 g/mm grating (blazed 
at 5000\AA\,).  The spectra were projected onto a 2048 $\times$ 515 CCD, 
with 1.0 arcsec/pixel plate scale perpendicular to the dispersion.  
The resulting spectra had an average resolving power of 3575, however the 
resolving power ranged from $\sim$ 3425 at the edges to $\sim$ 3685 at the 
center due to field curvature.  These values were determined by measuring 
the widths of emission line features in the spectra of Neon-Argon lamps.  
The wavelength coverage was slightly different for these runs. In February 
and May, the wavelength range was 6250-7500 \AA, while in August the wavelength range was 
shifted to 5950-7200 \AA\, to avoid the fringing that becomes problematic
($>$ few percent) longward of $\sim 7000$ \AA.  No extinction corrections
were applied to the spectra of these relatively close stars.

The images were reduced and the spectra were extracted using tools 
within the IRAF\footnote{IRAF is distributed by the National Optical 
Astronomy Observatories, which are operated by the Association of \
Universities for Research in Astronomy, Inc., under cooperative agreement 
with the National Science Foundation.} software package.  An average bias 
frame was subtracted from each flat field image, and these were then 
median combined to generate a normalized master flat for each night. All 
stellar spectra were likewise bias subtracted and then divided by the 
master flats.  The extracted spectra were background subtracted (using
a median fit) and the continuum of these extracted spectra were fit 
with a 7th order polynomial to produce normalized spectra.  The 
Neon-Argon spectra were extracted identically and were used to assign 
an approximate wavelength solution for each night. 

The observational epoch for each star is listed in Table 3 along with 
signal-to-noise ratio estimates for each spectrum.  The signal-to-noise values were
determined from the gain-corrected intensities in the center portions 
of the spectra.
The February run is not listed for any of the stars as only spectral 
standards were observed during this (poor weather) run.

\section{Spectroscopically Inferred Properties}\label{SIP}

For the 19 spectroscopically observed stars, measurements of the equivalent widths (EWs) of 
Li\,I 6708 \AA\, and H$\alpha$ are presented in Section \ref{EWLi}, and 
in Section \ref{MC} these measurements
are used to further assess the membership status of recently proposed 
members; none of these proposed members have been previously observed
at optical wavelengths.  In Sections \ref{FGK} and \ref{M} we describe 
our methods for determining the effective temperatures of these stars.

\subsection{Equivalent Widths of Li[I] and H$\alpha$}\label{EWLi}

Two common diagnostics of stellar youth are the atmospheric abundance 
of lithium and the amount of chromospheric activity. 
Of these two diagnostics, lithium is the least ambiguous because of its 
rapid depletion with age, which is largely attributed to proton capture 
\citep[e.g.][]{bod1965}.  The strength of the H$\alpha$ emission 
line is 
perhaps the most common tracer of chromospheric activity, but the 
persistence of H$\alpha$ emission, especially for the lowest mass stars,
makes it a much poorer diagnostic of age \citep{wes2008}.

\begin{figure}
\epsscale{.9}
\plotone{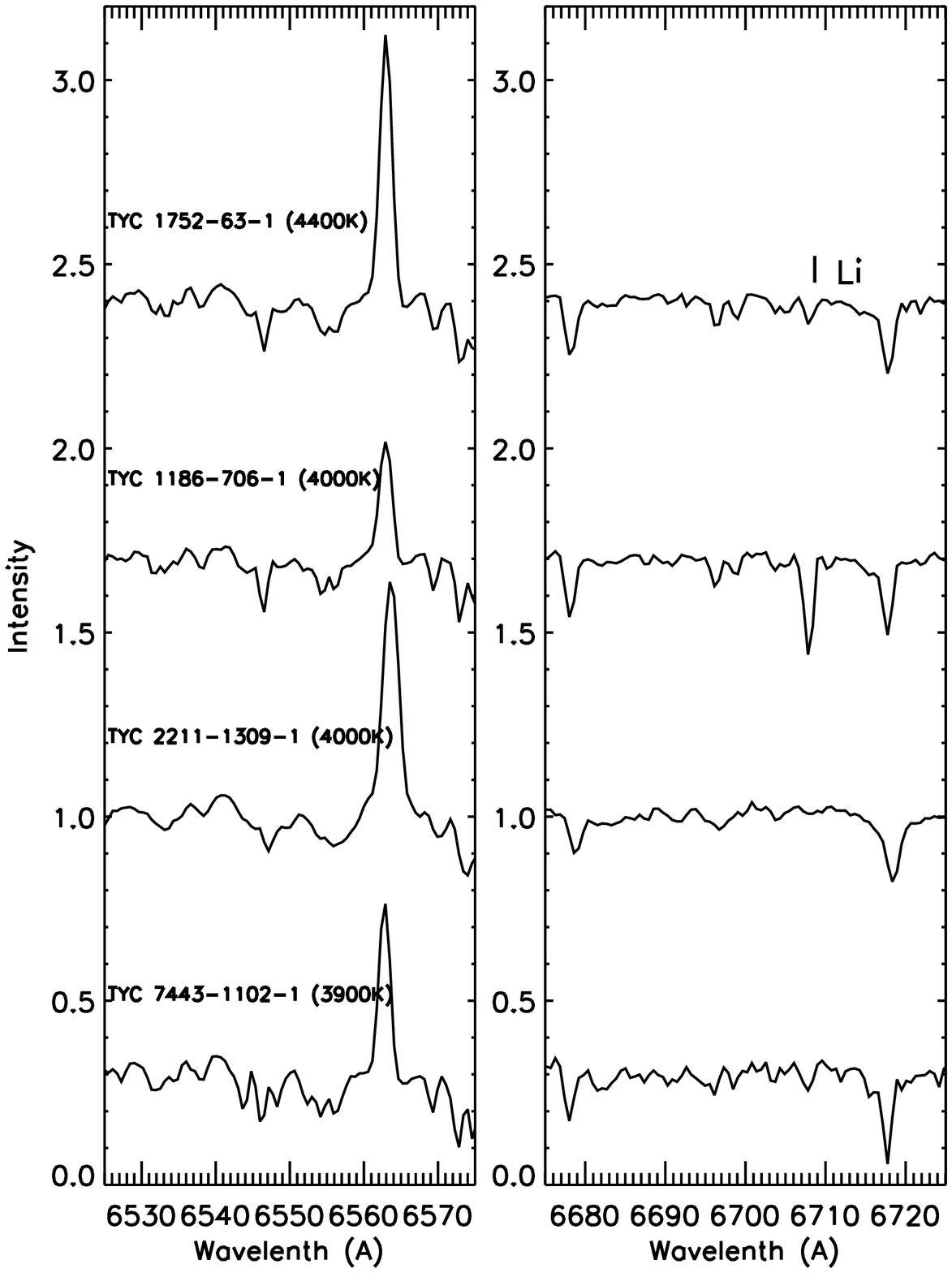}
\caption{Portions of DeVeny specrta of 4 Moving Group stars. The 
left panel shows portions near the H$\alpha$ emission feature (\emph{left panel}) and the Li\,I 6708 \AA\, absorption 
feature (\emph{right panel}). Labels give the names and determined temperatures. Lithium 
is detected in 3 of these 4 stars; the exception is the K7 star TYC 2211-1309-1.
\label{fig:Lithium}}
\end{figure}

Figure \ref{fig:Lithium} displays portions of DeVeny spectra near the H$\alpha$ and 
lithium 6708 \AA\, features. The EWs of 
Li\,I 6708 \AA\, were determined by fitting a Gaussian profile to the 
absorption feature, using tools within IRAF.  Because of the moderate 
resolution of these observations, the neighboring Fe\,I line at 6708 
\AA\, could not be removed from these measurements; 
the values may therefore be biased to larger EWs by 0.01 - 0.02 \AA.  
To measure the EW of the H$\alpha$ emission/absorption feature, a 
Gaussian profile was fit to this line after the deblending removal of 
the 5 nearby spectral features (telluric $H_{2}$O at 6552 \AA\,, 
Si[I] at 6555 \AA\,, Fe[I] at 6569 \AA\,, Ca[I] at 6572 \AA\,, and Fe[I] 
at 6574 \AA\,).  A negative sign indicates emission, following
convention.  The resulting measurements are listed in Table 
\ref{tab:LiandHa}.  The largest source of uncertainty in these measurements
is the determination of the continuum level, a somewhat subjective process. 
Therefore we assign uncertainties based on the EW variations over a range 
of reasonable continuum levels.  The EW uncertainties are on average 
0.02 \AA\, for the lithium measurements and 10\% for the H$\alpha$ 
measurements.

The measured EW values are listed in Table 3.  Of the 19 stars observed, 
11 exhibit H$\alpha$ in emission and 15 have detectable lithium absorption.
For the 4 stars without detectable lithium, detection upper limits are set 
by the continuum variations, and in most cases are 0.04 - 0.05 \AA.  The
one exception is 1RXS J19506.8-3320720, which has a low SNR spectrum and a 
much coarser lithium EW detection limit of $\sim 0.5$ \AA.  As a 
consistency check, we compare our Li 6708 \AA\, measurements to previous
measurements for the 13 stars with previously reported values.  With 1
exception, the values are consistent to within 0.03 \AA\, with the values 
reported by 
\citet[][3 stars]{men2008},
\citet[][3 stars]{wei2010}, 
\citet[][2 stars]{lop2006}, 
and \citet[][2 stars]{tor2006}.  The exception is LO Peg, for which
we measure an EW of $0.150\pm0.02$ \AA\, in a very high SNR spectrum, while 
\citet{lop2006} and \citet{sil2009} measure larger EW values of 
0.233 \AA\, and 0.215 \AA, respectively.

\subsection{Membership Confirmations}\label{MC}

Four stars observed here were recently proposed by \citet{lep2009} to be
members of the $\beta$ Pic Moving Group (TYC 1186-706-1, TYC 2211-1309-1, 
TYC 7443-1102-1 and 1RXS J19506.8-3320720).  All four exhibit H$\alpha$ 
emission, and
2 stars (TYC 1186-706-1, TYC 7443-1102-1) show clear Li\,I 6708 \AA\,
absorption, which confirms youth and greatly strengthens the case for 
association with this Moving Group.  Unfortunately the low signal-to-noise
ratio spectrum of 1RXS J19506.8-3320720 prevented us from determining
useful limits on the amount of lithium in its spectrum.  Lithium is not 
detected in the spectrum of TYC 2211-1309-1 (EW $< 0.04$ \AA), which
weakens the case for youth and association.  This star's K7 spectral 
type (determined below) is many hundreds of Kelvin hotter than the 
temperature of stars predicted to deplete their lithium fastest
\citep[e.g.][]{bar1998, yee2010}.  For example, the spectral type
M4 $\beta$ Pic members AT Mic A and AT Mic B exhibit no lithium in their 
spectra, but all high probability K7 members have strong Li\,I absorption
\citep[e.g. CP-72 2713: 0.44 \AA; CD-31 16041: 0.49 \AA;][]{sil2009}.
While this suggests TYC 2211-1309-1 is likely older than known
K-type members, it does not necessarily exclude it from the Moving
Group; there could be a age spread among members.

Only 1 other $\beta$ Pic star observed has no previously reported 
EW[Li] value, HD 15115, for which we measure a EW of $0.10 \pm 0.02$.  
However, with a spectral type of F8, this measurement does not 
significantly constrain the star's age, given the slow decrease in lithium 
abundance with time for F stars \citep{mam2008}.  We note that HD 15115
is the lowest probability member (at 60\%) listed in \citet{tor2008}.


TYC 1752-63-1 was recently proposed by \citet{sch2010} to be a member
of the AB Dor Moving Group.  This star exhibits H$\alpha$ emission and 
lithium absorption, both of which strengthen the case of an adolescent
age and association with the AB Dor Moving Group.

\subsection{F, G, and K Spectral Types}\label{FGK}
Measured ratios of temperature sensitive absorption lines in the spectra
of Sun-like stars are a powerful method of determining stellar effective 
temperature \citep[e.g.][]{str1990}.  In principle this is accomplished
best by measuring two absorption features of a particular element with substantially 
different excitation energies.  In practice, however, the success of this 
relies upon identifying features that can be measured precisely, given 
the spectral resolution and wavelength coverage, and then being able to
calibrate these ratios using spectra of stars with accurately determined 
temperatures.  To investigate the possibility of this with our spectra,
we searched by eye for absorption features within $\sim$ 10 \AA\, 
of each other that appear to vary inversely with temperature.  The 
temperature dependence is established from our DeVeny observations of 
stars in the study of \citet{val2005}, who determine effective temperatures 
for F-, G- and K-stars accurate to 44 K via comparisons with synthetic 
stellar spectra.  The EWs of these features are determined by fitting a 
Gaussian profile to the absorption feature using tools in IRAF.  
The 3 line pair ratios that show the strongest correlation between EW ratios and 
temperature and that we adopt for temperature determination in this study 
are 6137/6142 \AA, 6162/6170 \AA, and 6450/6456 \AA.  

To better understand the substratum of the temperature sensitivity of these line ratios,
we referred to the high resolution (R $\sim 350,000$) spectrum of the Sun
provided by \citet{lob2007}, who label solar absorption features and 
provide excitation energies.  From this we determine that the line ratio 
6137/6142 actually consist 
of Fe I 6136.615 (2.453 eV), Fe I 6136.993 (2.198 eV) and Fe I 6137.694 
(2.588 eV) versus Ba II 6141.713 (0.704 eV) and Fe I 6141.730 (3.603 eV); 
excitation energies are given in parentheses.  The line ratio 6162/6170
consists of Ca I 6162.173 (1.899 eV) versus Ca I 6169.042 (2.527 eV)
Ca I 6169.563 (2.526 eV) and Fe I 6170.504 (4.795 eV).  And finally, 
the line ratio 6450/6456 consists of Ca I 6449.808 (2.521 eV) versus
Ca I 6455.598 (2.523 eV) and Fe II 6456.383 (3.903 eV).  The temperature
sensitivity in these cases likely stems from the individual features
with the most discrepant excitation energies.  We also note that all
of these blends are unresolved in our modest resolution spectra as 
determined by the absence of any significant residuals of the best
fit Gaussians.

\begin{figure}
\epsscale{.9}
\plotone{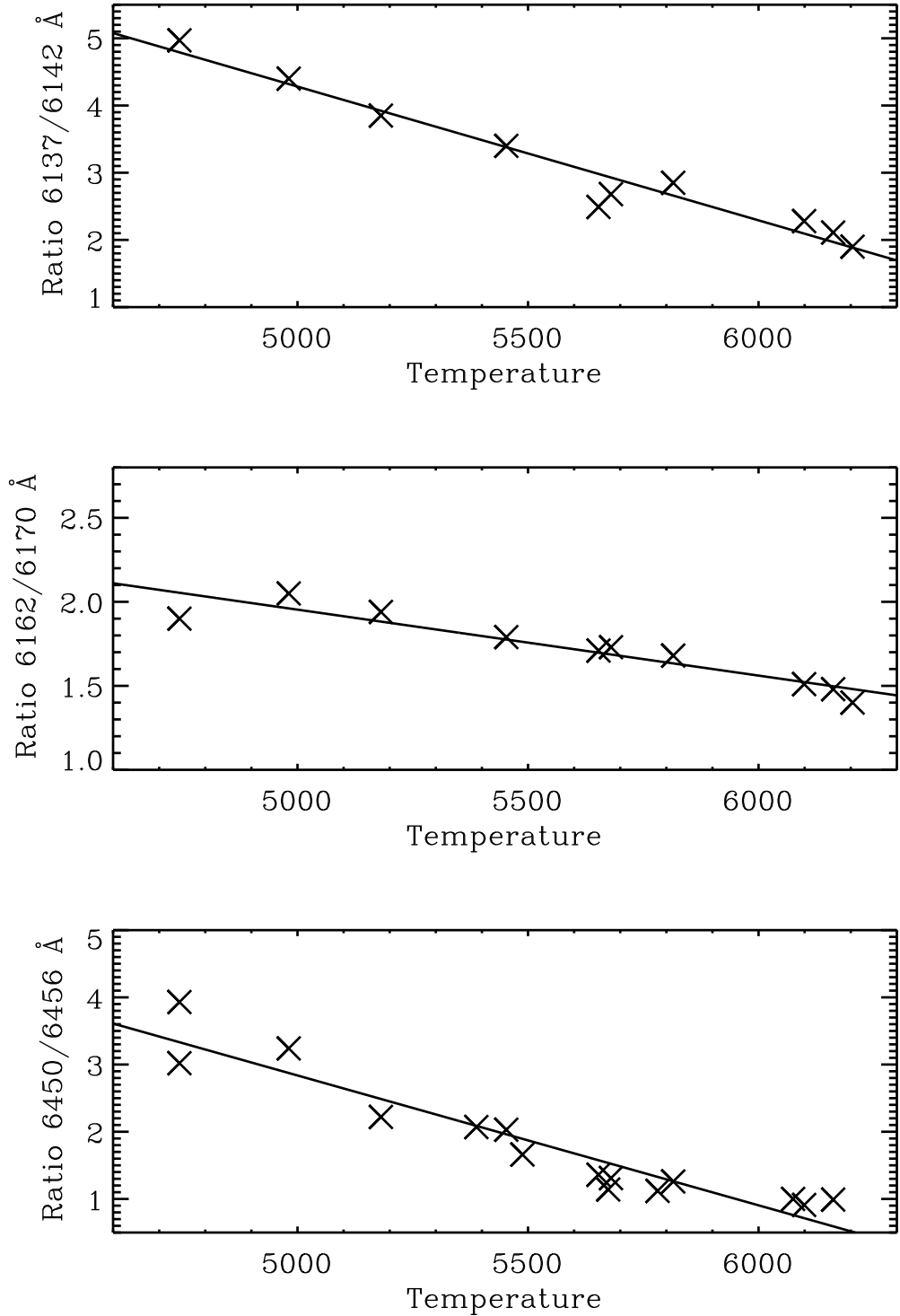}
\caption{Equivalent width ratios versus temperature for stars 
with temperatures determined by \citet{val2005}. The solid lines 
illustrate the best fit linear relations used to estimate 
temperatures for the observed Moving Group stars.  From top to bottom, the
dispersions about this best fit, a measure of uncertainty, are
114 K, 182 K, and 152 K, respectively. 
\label{fig:lineratio}}
\end{figure}

To quantify the temperature dependence of these line ratios, a linear
trend is fit to the line ratios versus temperature;
higher order fits appear unwarranted.  The best fit linear relations are 
shown in Figure \ref{fig:lineratio}.  Ten \citet{val2005} stars are
used to determine the best fit relation for the 6137/6142 and 6162/6170
ratios, while 15 stars are used to determine the 6450/6456 ratio; as noted
in Section 3, stars observed in the May run had a slightly redder wavelength
setting, and thus did not contain the two bluer line ratios.  
\citet{val2005} estimate metallicities for all of the stars
used in these calibrations, and although most have metallicity estimates 
within $\pm 0.12$ of solar, two stars are slightly metal rich, HD 182488 
([Fe/H] $= +0.22$) and HD 145675 ([Fe/H] $= +0.46$).  However, the the
best fit relations predict temperatures based on their measured ratios 
that are within 20 K of their independently determined values.  Thus,
there is no evidence that these relations are strongly metal 
dependent.


There are 2 primary sources of error inherent 
to this method for estimating temperature. The first 
stems from how well the standard stars agree with one another.  We
estimate this from the 1 $\sigma$ dispersion about the best fit linear 
relations illustrated in Figure \ref{fig:lineratio}, which are 114 K, 
182 K, and 152 K, respectively.  Since the temperature is determined by
equally weighting all 3 line ratio estimates, we estimate the combined 
error by averaging these 3 dispersion
estimates and dividing by $\sqrt{3}$.  This yields a typical uncertainty
of 86 K.  The second source of error stems from our ability to measure
the EWs of young stars due to the noise in the spectra. The typical 
EW uncertainty of $\pm$0.01 \AA\, corresponds to temperature uncertainty
of $\sim$ 60 K.  We add these 2 error terms in quadrature and assign 
temperature uncertainties of 105 K using this technique.

We used this method to estimate temperature estimates for the F-, G-,
and early-K stars ($T_{eff}$ ranging from 4600--6300 K)
in the observed sample, which is 11 of the 19 stars
observed.  For all but one star (PX Vir), the temperature assignments 
are an average of all 3 EW ratio estimates. Because PX Vir was 
observed only in May, the spectral range does not contain the features
from the first two ratios.  In this case, the composition of this 
star's temperature is solely determined from the 6450/6456 EW ratio, and
has a larger uncertainty of 163 K.  We also note that PX Vir is 
spectroscopic binary, and its companion star may introduce a systematic
error not accounted for in our temperature estimates.
The results of all 
temperature determinations are listed in Tables \ref{tab:ABDor} and 
\ref{tab:betapic}.

We compare these newly determined temperatures to temperatures inferred
from the previously assigned spectral types listed in Tables 1 and 2.
The previous spectral types 
originate from a variety of techniques, but most are determined via 
comparisons with optical spectra, similar to the method described in 
Section \ref{M}.  We assign temperatures
to these spectral types using the spectral type relationship in \citet{har1994}.
All but 2 stars have temperatures that are consistent to within 300 K,
or effectively 2 spectral subclasses.  The 2 exceptions are the F8 
stars HD 14082 AB and HD 15115, previously classified as F5 and F4, 
respectively.  We note however that HD 15115 is included in the 
\citet{val2005} spectral synthesis study, which we used to
calibrate our temperature relations \ref{FGK}, and they determine
a temperature of 6219 K, consistent with our determined temperature
of $6120\pm105$ K.  Our technique is at least self-consistent and
the apparent discrepancy may stem in part from the temperature/spectral 
type relation adopted for these comparisons.

\subsection{Late-K and M Spectral Types}\label{M}
 
For stars cool enough to be classified as late-K and M, their 
spectra transition from being dominated by atomic features to being
dominated by molecular band features which greatly inhibits the use of
atomic features to determine temperature.  For these stars we rely 
upon direct comparisons with stars of known spectral type and then 
adopt a spectral type - temperature scale to estimate temperatures.
 
\begin{figure}
\epsscale{.9}
\plotone{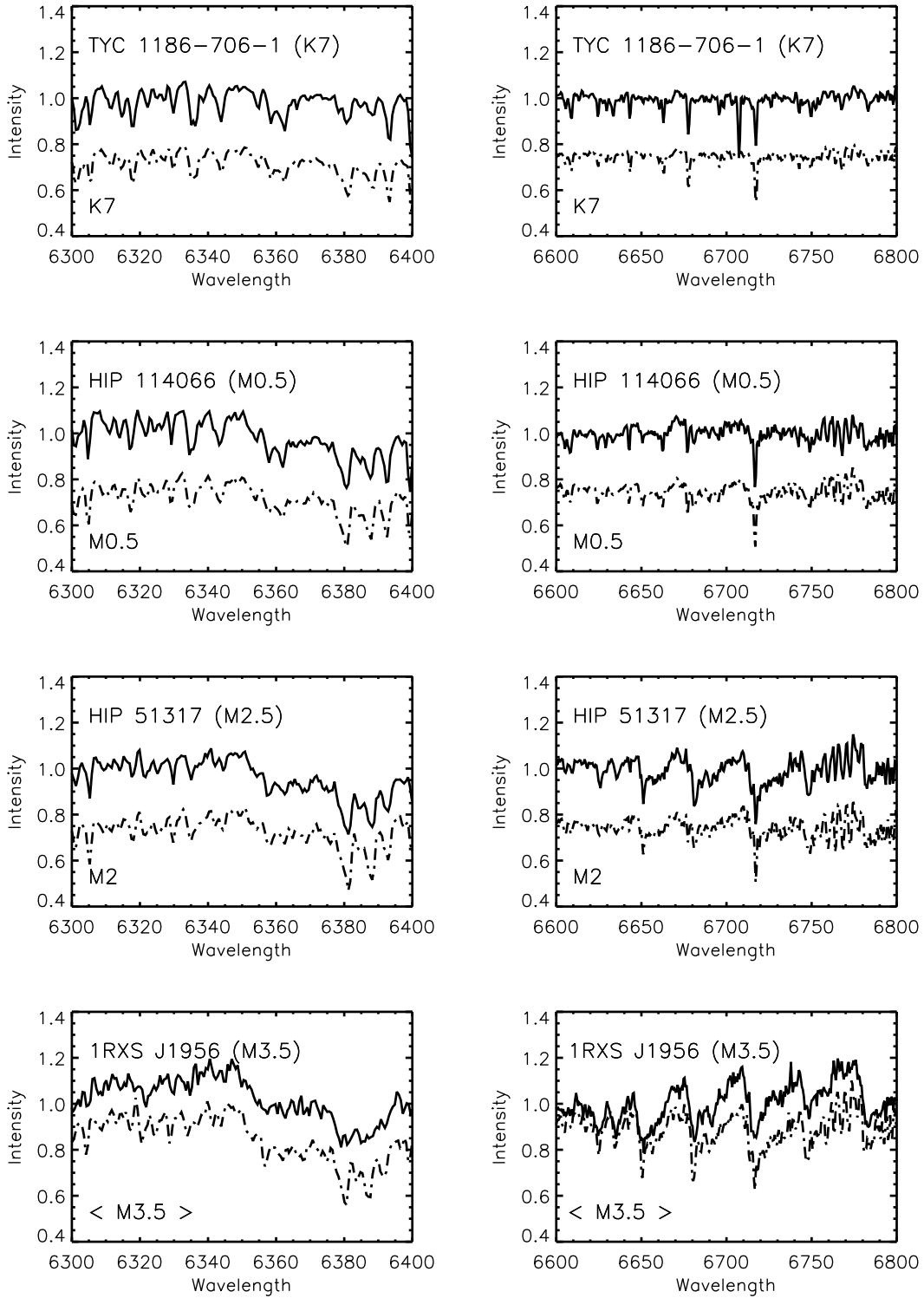}
\caption{Illustration of the method used to determine spectral types 
for late-K and -M stars. Spectral types are assigned to the observed 
spectra (\textit{solid lines}) based on the best matching spectral 
standard (\textit{dot-dashed lines}, labeled) or spectral standard 
average (labeled in brackets). 
\label{fig:MdwarfComparison}}
\end{figure}

Our spectral comparisons rely upon comparisons of newly obtained 
DeVeny spectra of stars classified as `primary standards' in 
\citet{kir1991}.  These types of comparisons are well known to yield 
spectral types accurate to at least 0.5 subclasses, given the strong 
variations of the TiO bands with temperature \citep[e.g.][]{web1999}.  
Because we were not able to obtain comparison standards at all M spectral 
subclasses, we approximate intermediate subclasses in some cases by 
averaging cooler and hotter spectra together (e.g. averaging an M3 and 
an M4 to approximate an M3.5).  Figure \ref{fig:MdwarfComparison} 
illustrates several examples of these comparisons for 4 Moving Group 
M dwarfs.  Based on the rapid changes in the spectra with temperature, 
spectral type uncertainties are 0.5 spectral subclasses.
 Out of the 19 observed Moving Group members, 8 spectral types are 
determined this way.  These newly assigned spectral types are consistent
with the ones assembled in \citet{tor2008} to 0.5 spectral subclasses,
with no systematic difference.

The effective temperatures assigned to the 8 late K and M dwarf stars
in this study are set using the spectral type/temperature scale from 
\citet{har1994}, which is appropriate for dwarf stars.  While there 
is evidence that very young ($\sim$ few Myr) M stars have a temperature
scale that is warmer than this, given their slightly extended atmosphere 
\citep{whi1999, luh2003}, there is no convincing evidence 
that this offset persists for many tens of millions of years.  The 
uncertainty of 0.5 spectral subclass corresponds to a temperature 
uncertainty $\sim$ 75 K.

\section{Angular Size Estimates}
 
In this section we estimate angular sizes for the 127 AB Dor 
and 77 $\beta$ Pic members assembled in Tables 1 and 2.  This
is accomplished by using the assembled temperatures along with
estimated luminosities to calculate stellar radii, that in
combination with the measured or estimated distances, can predict angular 
sizes.  The assembled temperatures and 
distances are described in Section 2.  Luminosities are 
determined by correcting the apparent $V$-band magnitudes to 
an absolute magnitude, and then to a bolometric luminosity using
bolometric corrections from \citet{har1994} for F-type stars and values from 
\citet{ken1995} for A and B stars. We adopt a solar bolometric 
luminosity of $M_{bol}(\odot)=4.83$ in these calculations 
\citep{glu2002}.  The bolometric corrections are done relative
to the $V$-band since the median spectral type of the sample is
late-G, which has an energy distribution that peaks at optical 
wavelengths.

Of the 204 stars listed as potential members of the AB Dor
and $\beta$ Pic Moving Groups, we are able to estimate angular
sizes for 167 stars (Tables 1 and 2).  The 37 stars for which we 
do not estimate sizes are in most cases companion stars that are 
too close to their primary to have spatially resolved photometry 
and/or a spectral type.  This close proximity makes them less ideal
targets for interferometric measurements in any case.
Twenty-five of the sizes estimated are of multiple star systems 
with spatially unresolved $V$-band magnitudes.  
This subset of size estimates, which are biased toward larger
values, are indicated in Tables 1 and 2 with brackets.
The estimated angular sizes range from 0.06 mas 
to 1.17 mas.  The 5 largest stars are the 
B6 star $\alpha$ Gru (1.17 mas), 
the M1 star AU Mic (0.72 mas), 
the A3 star $\beta$ Pic (0.69 mas), 
the M2.5 star GJ 393 (0.69 mas), 
and the F8 star HD 25457 (0.63 mas); 
all five are single stars.

As a check on our prescription for calculating angular sizes, we
apply the same methodology to predict the angular diameters of a sample of
stars whose diameters have been directly measured via interferometry.
Stars were selected from \citet{bel2009}, who report new measurements
from the Palomar Testbed Interferometer \citep{col1999} and assemble
previous CHARA Array measurements from \citet{bai2008}; both data sets
report angular sizes accurate to a few percent.  The comparison
sample includes late-B through early-M main sequence stars; we
use the spectral types, $V$ magnitudes, and distances provided in their
study.  The average difference between the measured and calculated
sizes is $-3$\%, although with a large dispersion of 21\%.  However,
many of the stars with the largest discrepancies
have temperatures assigned from spectral types that are vastly
different ($\ge$ 380 K) from the values inferred directly from interferometric
radius measurements.  The most extreme case is the HD 157881,
which has an interferometrically determined temperature of 3664
K, but an adopted K2 temperature of 4900 K.  In many of these
cases we suspect large spectral type errors; van Belle \& von
Braun (2009) did not redetermined spectral types, but assembled
them from uncited sources.  If we restrict
the sample to the 31 stars whose interferometrically determined
temperatures that agree with those from their spectral type to better
than 380 K (corresponding to approximately $\pm2$ spectral subclasses)
and include only dwarf stars, the average difference between the
measured and calculated sizes reduces to $-2$\% with a dispersion of 
8\%.  This result is also corroborated with a comparison of the 
subset of radii measurements that have uniformly determined spectral 
types from \citet{gra2001} and \citet{gra2003}.  For this subset
of 19 stars, the average difference between the measured and 
calculated sizes is $-5$\% with a dispersion of $10$\%.  Since the
spectral types of many of these newly identified Moving Group
members are determined from modern prescriptions, such as that
outlined in Sections 4.2 and 4.3, with temperature errors of less
than a few hundred Kelvin, we adopt 8\% as the typical size
uncertainty for the predicted sizes for stars with 
Hipparcos determined distances.

\section{Spatially Resolvable Young Stars}

\begin{figure}
\epsscale{1}
\plotone{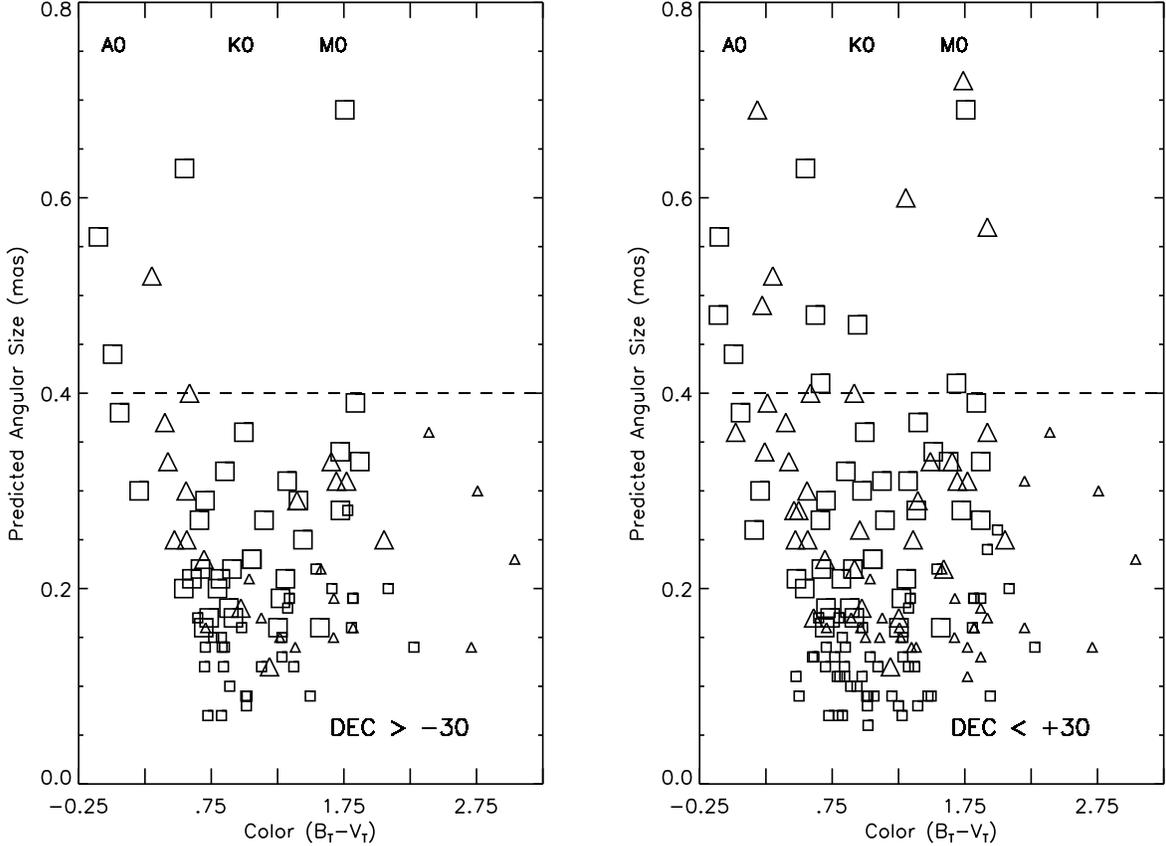}
\caption{Angular diameter versus Tycho $B_T - V_T$ color for AB Dor stars 
($squares$) and $\beta$ Pic stars ($triangles$).  The 
\textit{left panel} displays stars visible in the northern hemisphere 
(Dec $> -30$) while the \textit{right panel} shows stars visible from 
the south (Dec $< +30$).  One proposed member of AB Dor, $\alpha$ Gru
with a size of 1.17 mas is above the size range illustrated; at a declination of -47, it is
only accessible to southern facilities.
The \textit{dashed lines} represents a 
characteristic resolution limit for current long-baseline interferometric 
facilities.  \textit{Large symbols} represent infrared bright stars 
($K < 7.0$).  According to these criteria, there are 18 stars that are 
bright enough and large enough to observe in the southern hemisphere 
but only 6 such stars in the north (see Tables 1 and 2).
\label{fig:angsizetemp}}
\end{figure}

In Figure \ref{fig:angsizetemp} we illustrate the angular diameters
calculated for 167 of the 204 Moving Group members, all of which have $B_{T}-V_{T}$ 
color, $K$ magnitude, declination, and angular size estimates; one star 
($\alpha$ Gru, angular diameter = 1.17 mas) has a diameter larger than the range of this plot.
We distinguish subsets of this sample based on those that are likely 
large enough and bright enough to be spatially resolved by 
long-baseline optical/infrared interferometer.  We set a size of 0.4 
mas as the minimum size to be spatially resolved, consistent with what 
has been demonstrated with interferometric baselines larger than 300 m \citep[e.g.][]{bai2007}.
We use the $K_{2MASS}$ magnitudes to assess whether
the stars are bright enough to observe, adopting 7.0 as the practical
magnitude limit of operational interferometers.  However, of the 167 
stars for which we have size estimates, only 159 have $K_{2MASS}$ 
magnitudes; as noted in Section 2, fewer pairs have spatially resolved 
2MASS measurements than Tycho II measurements.  For these 8 stars
with missing 2MASS measurements, we estimate their $K_{2MASS}$ 
magnitudes using the resolved $V$ magnitude, spectral type, and the 
$V-K$ color relations of \citet{har1994} and \citet{ken1995}; we do not list these approximate $K$ 
magnitudes in the Tables.

Of the 167 Moving Group members with size estimates and 
$K_{2MASS}$ photometry, 18 are large enough ($> 0.4$ mas) and
bright enough ($K_{2MASS} < 7.0$) to be observed and spatially resolved.
However, only 6 stars are north enough (DEC $> -30$) to be easily 
accessible to northern interferometers, which currently has the longest
baseline interferometers.  These 6 include include the $\beta$ Pic stars 
51 Eri (F0; 0.52 mas) and AF Lep A (F7;
0.40 mas) and the AB Dor stars HD 17573 A (B8; 0.56 mas), HD 25457 (F8;
0.63 mas), GJ 393 (M2.5; 0.69 mas), and $\delta$ Scl (A0; 0.44 mas).
Of these, GJ 393 is of particular interest.
While the intermediate mass F, A and B stars should all be zero-age main 
sequence stars for even the youngest ages 
proposed for these associations, GJ 393, with a spectral type of M2.5, is not 
predicted to achieve this until an age of $\sim 100$ Myr 
(e.g. Baraffe et al. 1998; Siess et al. 2000).  Thus, 
depending upon the age of AB Dor, GJ 393 could be distinctively pre-main 
sequence and its location above the main sequence could help to constrain 
the age of this Moving Group.

Although all 18 stars are accessible to southern interferometers (DEC 
$< +30$), the shorter baselines of these facilities restrict the
sample resolvable to only the B6 star $\alpha$ Gru (1.17 mas).  
We nevertheless highlight that if resolutions comparable to 
CHARA are achievable, many stars could be spatially resolved since 
the majority of Moving Group members are in the southern hemisphere 
sky.  Two high-profile examples include the debris disk hosting M 
dwarf AU Mic (0.72 mas) and the young solar analogue AK Pic A (0.48 mas).
Size measurements of these stars would better constrain their ages and
the age (and possibly age spread) of the Moving Group.  Moreover,
these temporal stamps would help trace the rate of disk evolution
and possibly planetary system formation.

\section{Summary and Comments on Future Prospects}
 
We present moderate resolution (R $\sim$ 3575) optical spectra of 19 
members of the AB Dor and $\beta$ Pic Moving Groups, including 5 recently 
proposed members.  The strengths of H$\alpha$ emission and Li\,I 6708 
\AA\, 
absorption, both signatures of youth, are extracted from these spectra.
The detection of Li\,I 6708 \AA\, in the proposed $\beta$ Pic members TYC 
1186-706-1 (K7) and TYC 7443-1102-1 (K7.5) further strengthen the case
for youth and membership within this young Moving Group.  
No Li\,I is detected in the spectrum of the proposed $\beta$ Pic member TYC 2211-1309-1 
(K7).  Although this alone can
not refute its membership given the spread in Li\,I abundances for known
members \citep{men2008, sil2009}, it does suggest it is slightly older
than other known members; it would be the only $\beta$ Pic star of K 
spectral type without detected lithium.
Li\,I is also detected in the spectrum of the recently proposed member 
of the AB Dor Moving Group, TYC 1752-63-1 (K5), which likewise strengthens 
its case for membership.
In addition to these measurements, 
temperature sensitive line ratios are used to estimate
the temperatures of F through early-K stars, and direct comparisons with 
spectral standards are used to determine spectral types for late-K and 
M stars.

Updated membership lists for both Moving Groups are
assembled with an emphasis placed on identifying multiple star system
and spatially resolved photometric measurements.  Currently, the AB Dor moving group 
contains 127 proposed members and the $/beta$ Pic moving group holds 77 proposed members.
For these ensemble samples, temperatures determined from either new or
previous measurements are used in combination 
with distances and Tycho II photometry to predict angular sizes of 167
proposed members.
A comparison of sizes predicted in this way for a sample of main sequence 
stars that have been spatially resolved interferometrically implies that 
the size estimates are accurate to 8\%.
Six of these Moving Group members are bright enough ($K < 7.0$),
large enough ($\theta_{res} >$ 0.4 mas) and north enough (DEC $> -30$)
to be spatially resolvable with the CHARA Array, the world's longest
baseline interferometer working at optical/infrared wavelengths.
One of these stars is the low mass AB Dor member GJ 393; size
measurements of this likely pre-main sequence star could help
constrain the age of AB Dor, which is somewhat poorly determined.  
Using the
same brightness and resolution criteria, 18 stars could be observed
from the southern hemisphere (DEC $< +30$; includes all 6 northernly
accessible stars).  However, the operational baselines of southern
hemisphere interferometers currently will only allow the largest of
these, the B6 star $\alpha$ Gru (1.17 mas), to be spatially resolved.

Dramatically increasing the number of young stars that can be spatially 
resolved interferometrically would not only improve age estimates for 
these associations, but would also provide powerful constraints on the 
early evolution of young stars settling onto the main sequence, over a 
range of masses and with a range of rotational velocities.  This will
require improving the effective resolution limit of interferometers, since
the vast majority of young stars have sizes smaller than 0.4 mas (Figure 4).
Here we comment on the most practical way to achieve this with
an emphasis on the lowest mass stars that, according to
evolutionary models, should still pre-main sequence.

Interferometers are able to determine the angular size of a spatially 
resolved source by measuring its reduced fringe contrast, or visibility,
relative to that of an unresolved source.  For a perfectly calibrated 
system, unresolved sources have a visibility of 1.0 while resolved sources
have a visibility less than this \citep[see][for a review]{han2007}.  
Although the smallest angular size that an interferometer can measure is 
typically approximated as $\lambda$/(2$\times$baseline), which corresponds 
to 0.5 mas for H-band observations at a baseline of 331 m, in practice the 
true limit is set by how accurately the visibility can be measured; 
reductions in the visibility below 1.0 can also be caused by instrumental 
and/or environmental effects.  These effects are typically accounted for
by observing spatially unresolved stars to calibrate the system's 
visibility \citep[e.g.][]{bod2007}.  However, turbulence in the atmosphere
and vibrations along the optical path make these calibrations imperfect.  
Facilities that spatially filter the light from each telescope 
\citep[e.g.][]{cou1997} or that utilize adaptive optics systems on their 
telescopes \citep[e.g.][]{wiz2006} appear to successfully mitigate the 
former, while improved metrology systems may help mitigate the latter 
\citep[e.g.][]{wya2002}.  Meanwhile, more precisely determined visibilities could 
be obtained simply by greatly increasing the number of observations.
While we can not make specific recommendations on how to achieve this, 
we highlight that more robustly determined visibility measurements is 
one practical way of improving the effective resolution of current 
interferometers, thereby allowing more young stars to be spatially resolved.

A more direct way of increasing the number of stars that can be spatially 
resolved is to increase the baselines of either northern or southern 
facilities.  For example, doubling the longest baselines to $\sim 660$ m 
would improve the resolution to 0.2 mas, and allow 37 stars in the north 
and 65 stars in the south to be resolved (from Tables 1 and 2).  Moreover, 
this would permit a large fraction of stars in other Moving Groups (e.g. TW 
Hydrae) to also be resolved.  Although in principle this only requires 1 
additional telescope and in many cases is logistically feasible, this does 
raise an additional issue for visibility calibration.  As noted above, an 
important part of accurate size measurements is observations of unresolved 
stars (point sources) to calibrate the visibilities.
However, at baselines of 660 m, the majority of stars brighter than 
$K = 7$ are spatially resolved, making them somewhat poor stars for
visibility calibration.
Thus, to be fully realizable, this solution requires
not only the addition of telescopes to extend the baseline, but either
larger telescopes or a dramatic improvement in sensitivity that
permit observations of fainter, smaller stars that are spatially
unresolved.

An alternative to increasing the baseline to improve resolution is to
observe at shorter (e.g. optical) wavelengths.  Several facilities
already operate in the $V$, $R$, and $I$ bands, which offer resolution
improvements by factors of 2-4 relative to $H-$ or $K$-band.  If
we assume an optical limiting magnitude of $V=7$ and a resolution
limit of 0.2 mas, 20 AB Dor and $\beta$ Pic stars could be spatially
resolved, 9 accessible from northern facilities (DEC $> -30$) and 21
accessible from southern facilities (DEC $< +30$).  However, of these
20 stars, 16 have B, A or F spectral types; the 4 remaining stars are
classified as G (HD 45270, AK Pic A, V824 Ara A) or early K (AB Dor A) 
spectral type.  We note this because while
this option is likely to be realizable soon, it
will only permit Sun-like (or larger) stars to be observed which 
reach the main sequence in $\sim 30$ Myr or less.  Resolving a large
number of bona-fide pre-main sequence stars will require a sensitivity
to spectral type M stars, with a characteristic $V$ magnitude of 11.
This is again likely only achievable with 
larger telescopes or systems with dramatically improved sensitivity.




\acknowledgments

We are grateful for the observational assistance provided by Tom Bida 
and thank A. Reidel and T. Henry for their help in acquiring DeVeny 
spectra and for insightful comments during the inception of 
this project.  We also thank M. Simon and our astute referee for helpful comments on the first
version of this manuscript.
This project was funded by NSF/AAG grant \# 0908018 and a GSU Research 
Initiation Grant.

\clearpage



\clearpage

\clearpage


\clearpage


\clearpage

\clearpage

\begin{deluxetable}{rrrrrrrrrrrlr}
  \rotate
  \tabletypesize{\scriptsize}
  \tablewidth{0pt}
  \tablecaption{Proposed AB Doradus Moving Group Members\label{tab:ABDor}}
  \tablehead{
    \multicolumn{3}{c}{} &
    \colhead{Dist$^{a}$} & 
    \colhead{$V^{b}$} & 
    \colhead{$K^{b}$} &
    \colhead{Prev.$^{c}$} & 
    \colhead{$T_\textrm{eff}$} & 
    \colhead{New} & 
    \colhead{Lum$^{b}$} & 
    \colhead{$\theta^{b}$} & 
    \colhead{Known$^{e}$} & 
    \colhead{Mult$^{f}$} \\
    \colhead{HD} & 
    \colhead{HIP}  &
    \colhead{Other} &
    \colhead{(pc)} & 
    \colhead{mag} &
    \colhead{mag} & 
    \colhead{SpT} &               
    \colhead{(K)} & 
    \colhead{SpT,$T_\textrm{eff}$} &  
    \colhead{($L_{\odot}$)} & 
    \colhead{(mas)} & 
    \colhead{Mult} & 
    \colhead{Ref}  }  \startdata
      1405  &  \nodata  &  PW And            & 27    & 8.859 & 6.39  &  K2 & 4900  &  \nodata  & 0.24  & 0.23  & S     & L07,NC10,E11 \\
      \nodata  &  \nodata  &  TYC 1741-2117-1   & 47.6  & 11.167 & 7.66  &  K7    & 4000  &  \nodata  & 0.16  & 0.16  & \nodata  & \nodata \\
      4277 A & 3589 A &  BD+54 144 A          & 48.6H & [7.803] & [6.36] &  F8    & 6190  &  \nodata  & [1.6]  & [0.21] & B     & E11 \\
    B     & B     &  BD+54 144 B      & 48.6* & \nodata & \nodata &  \nodata  &  \nodata  &  \nodata  & \nodata & \nodata & B 3\farcs8 & T08 \\
    6569  & 5191  &  BD-15 200            & 50.0H & 9.446 & 7.34  &  K1    & 5080  &  \nodata  & 0.45  & 0.16  & S & E11 \\
      \nodata  & 6276  &  BD-12 243            & 35.1H & 8.391 & 6.55  &  G9    & 5410  &  \nodata  & 0.53  & 0.22  & S  & E11 \\
      \nodata  &  \nodata  &  TYC 1752-63-1     & 38.5  & 10.723 & 7.64  &  K5    & 4395  &  K5,4400  & 0.11  & 0.14  & \nodata  & \nodata \\
      \nodata  & \nodata &  CD-46 644 A & 70    & 11.136 & [8.61] &  K3  & 4730  &  \nodata  & 0.21  & 0.09  & B     & \nodata \\
      \nodata & \nodata &  CD-46 644 B & 70*   & \nodata & \nodata & \nodata & \nodata & \nodata & \nodata & \nodata & B 21\farcs4 & 2M \\
    13482 & 10272 &  BD+23 296 A & 32.3H & 7.932 & [5.73] &  K1    & 5090  &  \nodata  & 0.74  & 0.32  & B     & \nodata \\
      B & B &  BD+23 296 B & 32.3* & 9.519 & \nodata & K4:   & 4520  &  \nodata  & 0.20  & 0.21  & B 1\farcs8 & T08 \\
      16760 Aa & 12638 Aa &  BD+37 604 Aa      & 50.2H & [8.764] & [7.03]  &  G2    & 5860  &  \nodata  & [0.70]  & [0.15]  & T     & E11 \\
      16760 Ab & 12638 Ab &  BD+37 604 Ab      & 50.2* & \nodata & \nodata & \nodata& \nodata & \nodata & \nodata & \nodata  & T 0\farcs3  & E11 \\
            B & 12635 B &  BD+37 604 B         & 44.9H & 10.270 & 7.76  &  K2    & 4900  &  \nodata  & 0.18  & 0.12  & T 14\farcs6 & 2M,E11 \\
      17332 A & 13027 A &  BD+18 347 A         & 32.6H & 7.446 & 5.52  &  G1    & 5940  &  \nodata  & 1.0  & 0.27  & B     & E11 \\
            B & B     &  BD+18 347 B       & 32.6* & 8.170 & 5.64  &  G6    & 5690  &  \nodata  & 0.55  & 0.21  & B 3\farcs6 & T08 \\
      17573 A & 13209 A &  BD+26 471 A         & 48.9H & [3.594] & [3.86] &  B8     & 11900 &  \nodata  & [160]   & [0.56] & B     & \nodata \\
            B & B     &  BD+26 471 B    & 48.9* & \nodata & \nodata &  \nodata  &  \nodata  &  \nodata  &  \nodata  &  \nodata  & B 0\farcs2 & M90 \\
    19668 & 14684 &  IS Eri              & 40.2H & 8.492 & 6.70  &  G0    & 6030  &  G5,5760  & 0.59  & 0.18  & S  & E11 \\
    \nodata & 14809 &  BD+21 418 A         & 49.4H & 8.513 & 6.97  &  G5    & 5750  &  G1,5970  & 0.87  & 0.16  & T     & E11 \\
    \nodata & 14807 A & BD+21 418 Ba         & 51.7H & [10.463] & [7.65]  &  K6     & 4260  &  \nodata  & [0.32]  & [0.18]  & T 33\farcs1 & 2M,E11 \\
    \nodata & 14807 B & BD+21 418 Bb        & 51.7* & \nodata & \nodata  & \nodata & \nodata & \nodata & \nodata & \nodata  & T 0\farcs3 & 2M,E11 \\
    \nodata & \nodata & HW Cet & 46c   & 10.270 & 8.02  & K4:   & 4520  &  \nodata  & 0.21  & 0.15  & \nodata  & \nodata \\
    20888 & 15353 &  TYC 8866-1472-1     & 58.0H & 6.022 & 5.69  &  A3     & 8720  &  \nodata  & 13  & 0.26  & \nodata  & \nodata \\
      21845 A & 16563 A &  V577 Per A          & 33.8H & 8.250 & 6.37  &  G5    & 5750  &  \nodata  & 0.53  & 0.20  & B     & NC10,E11 \\
            B & B     &  V577 Per B        & 33.8* & 11.188 & 7.59  &  M0    & 3800  &  \nodata  & 0.10  & 0.20  & B 9\farcs5 & 2M,E11 \\
      \nodata  & 17695 &  RX J0347.3-0158     & 16.3H & 11.488 & 6.93  &  M3   & 3350  &  \nodata  & 0.036  & 0.33  & S     & L07,NC10,E11 \\
    24681 &  \nodata  &  BD-02 754           & 53    & 9.042 & 7.25  &  G8    & 5570  &  \nodata  & 0.65  & 0.15  & \nodata  & \nodata \\
    25457 & 18859 &  BD-00 632             & 19.2H & 5.378 & 4.18  &  F6   & 6350  &  F8,6200  & 2.2  & 0.63  & S     & L07,NC10,E11 \\
    25953 & 19183 &  BD+01 699             & 55.3H & 7.825 & 6.58  &  F5     & 6440  &  F7,6250  & 1.9  & 0.20  & S & E11 \\
      \nodata  &  \nodata  &  TYC 0091-0082-1   & 37c   & 10.931 & 8.65  &  K0    & 5250  &   \nodata   & 0.059  & 0.07  & \nodata & \nodata \\
      \nodata  &  \nodata  &  TYC 5899-26-1     & 16    & 11.632 & 6.89  &  M3   & 3350  &  \nodata  & 0.032  & 0.31  & \nodata & \nodata \\
      \nodata  & 22738 A &  GJ 2036 A                 & 11.2H & 10.991 & 6.34  &  M3   & 3350  &  \nodata  & 0.027  & 0.41  & B     & \nodata \\
      \nodata  & B     &  GJ 2036 B               & 11.2* & 11.896 & 6.89  &  M3   & 3350  &  \nodata  & 0.012  & 0.27  & B 7\farcs7 & 2M \\
      31652 &  \nodata  &  BD-09 1034       & 88    & 9.981 & 8.35  &  G8    & 5570  &  \nodata  & 0.75  & 0.10  & \nodata & \nodata \\
      \nodata  &  \nodata  &  CD-40 1701        & 42    & 10.564 & 8.05  &  K4    & 4590  &  \nodata  & 0.13  & 0.13  & \nodata & \nodata \\
    32981 &  \nodata  &  BD-16 1042          & 81    & 9.119 & 7.74  &  F9    & 6120  &  \nodata  & 1.3  & 0.12  & \nodata & \nodata \\
    293857 &  \nodata  &  BD-04 1063         & 78    & 9.244 & 7.36  &  G8    & 5570  &  \nodata  & 1.2  & 0.14  & \nodata & \nodata \\
    33999 A &  \nodata  &  CD-34 2128 A        & 106   & [8.858] & [7.20] &  F8    & 6190  &  \nodata  & [2.8]  & [0.13] & T     & \nodata \\
    B     &  \nodata  &  CD-34 2128 B      & 106*  & \nodata & \nodata & \nodata  &  \nodata  &  \nodata  &  \nodata  &  \nodata  & T 0\farcs7 & T08 \\
    C     & \nodata &  CD-34 2128 C & 106*  & \nodata & \nodata & \nodata &  \nodata  &  \nodata  &  \nodata  &  \nodata  & T SB & T08 \\
    35650 & 25283 &  CD-39 1951           & 17.7H & 9.144 & 5.92  &  K6    & 4260  &  \nodata  & 0.13  & 0.34  & \nodata  & \nodata \\
    36705 A & 25647 &  AB Dor A & 14.9H & [6.938] & [4.69] &  K0   & 5250  &  \nodata  & [0.38]  & [0.47] & Q     & \nodata \\
    C     &  C  &  AB Dor C & 14.9* & \nodata & \nodata &  \nodata  &  \nodata  &  \nodata  &  \nodata  &  \nodata  & Q 0\farcs2 & T08 \\
               Ba &  Ba  &  AB Dor Ba & 14.9* & [12.70] & \nodata &  M4    & 3150  &  \nodata  & [0.013]  & [0.24] & Q 9\farcs2 & 2M \\
    Bb    &  Bb  &  AB Dor Bb & 14.9* & \nodata & \nodata &  \nodata  &  \nodata  &  \nodata  &  \nodata  &  \nodata  & Q 0\farcs06 & T08 \\
      \nodata  & \nodata &  UX Col                    & 57    & 10.589 & 7.76  &  K3   & 4730  &  \nodata  & 0.23  & 0.12  & \nodata  & \nodata \\
      \nodata  & \nodata &  CD-34 2331                & 78    & 11.835 & 9.12  &  K3   & 4730  &  \nodata  & 0.14  & 0.07  & \nodata  & \nodata \\
    37572 A & 26373 A &  UY Pic A & 23.9H & 7.909 & 5.81  &  K0    & 5250  &  \nodata  & 0.40  & 0.30  & B     & NC10 \\
    B     & 26369 B &  UY Pic B & 24.3H & 9.571 & 6.61  &  K6  & 4260  &  \nodata  & 0.15  & 0.28  & B 18\farcs3 & 2M \\
      37551 A & 26401 &  WX Col A            & 74.7H & 9.674 & 7.66  &  G7    & 5620  &  \nodata  & 0.72  & 0.11  & B     & \nodata \\
            B & B &  WX Col B            & 74.7* & 10.600 & 7.87  &  K1  & 5080  &  \nodata  & 0.35  & 0.09  & B 3\farcs9 & T08 \\
      \nodata  & \nodata &  Parenago 2752       & 116   & 10.830 & 9.14  &  G8   & 5570  &  \nodata  & 0.60  & 0.07  & \nodata & \nodata \\
      \nodata  & \nodata &  CPD-19 878          & 71    & 10.710 & 8.12  &  K1    & 5080  &  \nodata  & 0.28  & 0.09  & \nodata & \nodata \\
      \nodata  & \nodata &  TYC 7605-1429-1 & 128   & 12.281 & 9.12  &  K4  & 4590  &  \nodata  & 0.25  & 0.06  & \nodata & \nodata \\
      \nodata  & \nodata &  CD-26 2425          & 70    & 10.870 & 8.47  &  K2   & 4900  &  \nodata  & 0.25  & 0.09  & \nodata & \nodata \\
    39576 & 27727 &  TZ Col                & 87.6H & 9.034 & 7.52  &  G3    & 5820  &  \nodata  & 1.7  & 0.14  & \nodata & \nodata \\
      \nodata  & \nodata &  TY Col              & 68    & 9.527 & 7.63  &  G6 & 5690  &  \nodata  & 0.66  & 0.11  & \nodata & \nodata \\
      \nodata  & \nodata &  BD-13 1328          & 39    & 10.587 & 7.77  &  K4 & 4590  &  \nodata  & 0.11  & 0.13  & \nodata & \nodata \\
      \nodata  & \nodata &  CD-34 2676          & 72    & 10.160 & 8.20  &  G9   & 5410  &  \nodata  & 0.44  & 0.10  & \nodata & \nodata \\
      \nodata  & \nodata & CD-35 2722 A & 24    & 10.954 & [7.05] &  M1   & 3650  &  \nodata  & 0.074  & 0.26  & B     & \nodata \\
      \nodata  &   \nodata  & CD-35 2722 B & 24    & \nodata & \nodata & L4    &  \nodata  &  \nodata  &  \nodata  &  \nodata  & B 0\farcs07 & W11 \\
    45270 & 30314 &  CD-60 1425 & 23.5H & 6.517 & [5.05] &  G1    & 5940  &  \nodata  & 1.3  & 0.41  & S     & NC10 \\
      \nodata  & \nodata &  GSC 8894-0426       & 24    & 12.7  & 7.21  &  M3   & 3350  &  \nodata  & 0.027  & 0.19  & \nodata & \nodata \\
      48189 A & 31711 &  AK Pic A            & 21.7H & 6.257 & [4.54] &  G2    & 5860  &  \nodata  & 1.4  & 0.48  & B     & NC10 \\
             B &  B  &  AK Pic B        & 21.7* & 8.639 & \nodata & K5:   & 4400  &  \nodata  & 0.25  & 0.37  & B 0\farcs8 & T08 \\
      \nodata & 31878 &  CD-61 1439         & 21.9H & 9.682 & 6.50  &  K7 & 4000  &  \nodata  & 0.14  & 0.33  & \nodata & \nodata \\
      \nodata  & \nodata &  TYC 7627-2190-1     & 78    & 11.085 & 8.81  &  K2   & 4900  &  \nodata  & 0.25  & 0.08  & \nodata & \nodata \\
      \nodata  & \nodata &  GSC 8544-1037       & 143   & 11.5  & 8.93  &  K4    & 4590  &  \nodata  & 0.65  & 0.08  & \nodata & \nodata \\
      \nodata  & \nodata &  CD-57 1654          & 103   & 10.438 & 8.93  &  G2    & 5860  &  \nodata  & 0.63  & 0.07  & \nodata & \nodata \\
      \nodata  & \nodata &  BD+20 1790          & 26    & 10.022 & 6.88  &  K5  & 4400  &  \nodata  & 0.10  & 0.19  & S  & E11 \\
    59169 A & 36108 &  CD-49 2843 A & 118H  & 10.215 & [7.93] &  G7    & 5620  &  \nodata  & 1.1  & 0.09  & B     & \nodata \\
    B     & B &  CD-49 2843 B & 118*  & 11.148 & \nodata & K3:   & 4700  &  \nodata  & 0.59  & 0.09  & B 1\farcs2 & T08 \\
      \nodata  & 36349 A &  V372 Pup A       & 15.6H & [10.077] & [5.72] &  M1   & 3650  &  \nodata  & [0.073]  & [0.39] & T     & \nodata \\
      \nodata & B     &  V372 Pup B & 15.6* &  \nodata  & \nodata & \nodata &  \nodata  &  \nodata  &  \nodata  &  \nodata  & T 0\farcs2 & T08 \\
      \nodata & C     &  V372 Pup C & 15.6* & \nodata & \nodata & \nodata &  \nodata  &  \nodata  &  \nodata  &  \nodata  & T 6\farcs6 & T08 \\
      \nodata  & \nodata &  CD-84 80            & 71    & 9.961 & 7.91  &  G9    & 5410  &  \nodata  & 0.51  & 0.11  & \nodata & \nodata \\
    64982 A & 37855 A &  CD-79 300 A          & 83.5H & 8.955 & [7.59] &  G0    & 6030  &  \nodata  & 1.6  & 0.13  & B     & \nodata \\
    B     & B     &  CD-79 300 B      & 83.5* & \nodata & 8.88  &  \nodata  &  \nodata  &  \nodata  &  \nodata  &  \nodata  & B 5\farcs7 & 2M \\
      \nodata  & \nodata &  BD-07 2388          & 93    & 9.373 & 6.92  &  K1 & 5080  &  \nodata  & 1.7  & 0.17  & \nodata & \nodata \\
    82879 & \nodata  &  RX J0928.5-7815 & 58c   & 8.987 & 7.83  &  F4    & 6580  &   \nodata   & 0.74  & 0.11  & \nodata & \nodata \\
      \nodata  & \nodata &  CD-45 5772          & 70    & 10.888 & 8.07  &  K4    & 4590  &  \nodata  & 0.19  & 0.09  & \nodata & \nodata \\
      \nodata  & 51317 &  GJ 393              & 7.23H & 9.586 & 5.31  &  M2    & 3500  &  M2.5,3420  & 0.035  & 0.69  & S     & L07,NC10,E11 \\
    99827 A & 55746 A &  CD-84 114 A           & 82.9H & [7.638] & [6.49] &  F5     & 6440  &  \nodata  & [5.2]  & [0.21] & B     & \nodata \\
    B     & B     &  CD-84 114 B & 82.9* & \nodata & \nodata & \nodata &  \nodata  &  \nodata  &  \nodata  &  \nodata  & B 3\farcs5 & T08 \\
      \nodata  & \nodata &  TYC 4943-192-1      & 45.5  & 11.344 & 7.83  &  M0    & 3800  &  \nodata  & 0.16  & 0.19  & \nodata & \nodata \\
    113449 A & 63742 A &  PX Vir A         & 22.1H & [7.700] & [5.51] &  K1    & 5080  &  K1,5020  & [0.43]  & [0.36] & B     & \\
    B     & B     &  PX Vir B & 22.1* & \nodata & \nodata & \nodata &  \nodata  &  \nodata  &  \nodata  &  \nodata  & 0\farcs4 & T08,NC10,E11 \\
    139751 A & 76768 &  BD-18 4125 A         & 42.6H & [10.377] & [6.95] &  K5   & 4400  &  \nodata  & [0.19]  & [0.16] & B     & \nodata \\
    B     & \nodata  &  BD-18 4125 B      & 42.6* & \nodata & \nodata &  \nodata  &  \nodata  &  \nodata  & \nodata & \nodata & B 0\farcs9 & T08 \\
      \nodata  & 81084 &  LP 745-70          & 31.9H & 11.178 & 7.55  &  M0    & 3800  &  \nodata  & 0.090  & 0.20  & S     & L07,NC10,E11 \\
    152555 & 82688 &  BD-04 4194           & 47.6H & 7.818 & 6.36  &  G0     & 6030  &  G0,6000  & 1.5  & 0.22  & S  & E11 \\
    317617 & \nodata  &  TYC 7379-279-1     & 56    & 10.452 & 7.67  &  K3    & 4730  &  \nodata  & 0.25  & 0.12  & \nodata & \nodata \\
    159911 & \nodata  &  BD-13 4687         & 45    & 10.100 & 6.84  &  K4   & 4590  &  \nodata  & 0.23  & 0.16  & \nodata & \nodata \\
      160934 A & 86346 &  GJ 4020 A          & 24.5H & [10.150] & [6.81] &  K7   & 4000  &  K7.5,3900   & [0.12]  & [0.29] & T     & NC10,E11 \\
             B & \nodata  &  GJ 4020 B & 24.5* & \nodata & \nodata & \nodata  &  \nodata  &  \nodata  &  \nodata  &  \nodata  & T 0\farcs2 & T08,NC10,E11 \\
    C     &   \nodata  &  GJ 4020 C & 24.5* & \nodata & 9.42  & \nodata &  \nodata  &  \nodata  &  \nodata  &  \nodata  & T 19\farcs1 & 2M \\
    176367 A & 93375 A &  CD-28 15269 A & 62.8H & 8.469 & 7.15  &  G1   & 5940  &  \nodata  & 1.4  & 0.17  & B     & \nodata \\
    B     & B     &  CD-28 15269 B & 62.8* & \nodata & 12.72 & \nodata &  \nodata  &  \nodata  &  \nodata  &  \nodata  & B 11\farcs2 & 2M \\
    177178 & 93580 &  BD+01 3865 & 55.0H & 5.815 & 5.32  &  A4     & 8460  &  \nodata  & 14  & 0.30  & \nodata & \nodata \\
    178085 & 94235 &  CD-60 7126           & 57.2H & 8.308 & 6.88  &  G1   & 5940  &  \nodata  & 1.4  & 0.18  & \nodata & \nodata \\
    181869 & 95347 &  $\alpha$ Sgr         & 52.1H & 3.943 & 4.20  &  B8     & 11900 &  \nodata  & 130   & 0.48  & \nodata & \nodata \\
      \nodata & \nodata &  TYC 486-4943-1    & 71    & 11.283 & 8.66  &  K3    & 4730  &  \nodata  & 0.19  & 0.08  & \nodata & \nodata \\
    189285 & \nodata  &  BD-04 4987          & 95    & 9.433 & 7.84  &  G7    & 5620  &  \nodata  & 1.4  & 0.12  & \nodata & \nodata \\
      \nodata & \nodata  &  BD-03 4778          & 70    & 10.011 & 7.92  &  K1   & 5080  &  \nodata  & 0.52  & 0.12  & \nodata & \nodata \\
      \nodata & \nodata  &  BD+05 4576          & 38.5  & 10.518 & 7.13  &  K7   & 4000  &  \nodata  & 0.19  & 0.22  & \nodata & \nodata \\
    199058 & \nodata  &  BD+08 4561          & 75    & 8.614 & 6.97  &  G5    & 5750  &  G6,5710  & 1.9  & 0.17  & \nodata & \nodata \\
      \nodata & \nodata  &  TYC 1090-543-1   & 75    & 11.284 & 8.82  &  K4   & 4590  &  \nodata  & 0.22  & 0.09  & \nodata & \nodata \\
    201919 & \nodata  &  GSC 06351-00286     & 39    & 10.433 & 7.58  &  K6   & 4260  &  \nodata  & 0.18  & 0.19  & \nodata & \nodata \\
      \nodata & 106231 &  LO Peg              & 25.1H & 9.245 & 6.38  &  K5   & 4400  &  K5,4450  & 0.19  & 0.27  & S & NC10,E11 \\
    207278 & 107684 &  CD-40 14502         & 83.7H & 9.614 & 7.98  &  G7    & 5620  &  \nodata  & 0.95  & 0.11  & \nodata & \nodata \\
      \nodata & 107948 &  GJ 4231             & 31.8H & 12.040 & 7.39  &  M2   & 3500  &  \nodata  & 0.059  & 0.19  & \nodata & \nodata \\
    209952 & 109268 &  $\alpha$ Gru        & 31.1H & 1.7   & 2.02  &  B6     & 14000 &  \nodata  & 520   & 1.17  & \nodata & \nodata \\
      \nodata  & 110526 A &  GJ 856 A           & 15.5H & 11.420 & [6.05] &  M3    & 3350  &  \nodata  & 0.038  & 0.34  & B     & E11 \\
      \nodata  & B     &  GJ 856 B           & 15.5* & 11.830 & \nodata &  M3    & 3350  &  \nodata  & 0.026  & 0.28  & B 1\farcs8 & T08 \\
    217343 & 113579 &  CD-26 16415         & 32.0H & 7.484 & 5.94  &  G5    & 5750  &  \nodata  & 0.95  & 0.29  & S & E11 \\
    217379 A & 113597 A &  CD-26 16420 A       & 30.0H & [10.024] & [6.27] &  K7    & 4000  &  \nodata  & [0.19]  & [0.28] & T     & \nodata \\
    B     & B     &  CD-26 16420 B & 30.0* & [10.620] & \nodata & \nodata  &  \nodata  &  \nodata  &  \nodata  &  \nodata  & T 1\farcs8 & T08 \\
    C     & C     &  CD-26 16420 C & 30.0* & \nodata & \nodata & \nodata &  \nodata  &  \nodata  &  \nodata  &  \nodata  & T SB & T08 \\
      \nodata & 114066 &  GJ 9809             & 24.9H & 10.922 & 6.98  &  M1    & 3650  &  M0.5,3720  & 0.076  & 0.25  & S  & NC10,E11 \\
      218860 A & 114530 &  CD-45 14955 A     & 50.6H & 8.803 & 7.03  &  G8    & 5570  &  \nodata  & 0.75  & 0.17  & B     & \nodata \\
             B &  B &  CD-45 14955 B     & 50.6* & 13.0  & 8.85  &  M3   & 3350  &  \nodata  & 0.091  & 0.16  & B 19\farcs6 & 2M \\
      \nodata  & 115162 &  BD+41 4749         & 49.4H & 8.929 & 7.22  &  G4     & 5780  &  \nodata  & 0.58  & 0.14  & S & E11 \\
    220825 & 115738 &  $\kappa$ Psc        & 49.7H & 4.917 & 4.90  &  A0     & 9520  &  \nodata  & 30 & 0.38  & \nodata & \nodata \\
    222575 & 116910 &  CD-36 15990         & 62.3H & 9.382 & 7.62  &  G8   & 5570  &  \nodata  & 0.65  & 0.13  & \nodata & \nodata \\
    223352 & 117452 &  $\delta$ Scl        & 44H   & 4.6   & 4.53  &  A0     & 9520  &  \nodata  & 32 & 0.44  & \nodata & \nodata \\
    224228 & 118008 &  GJ 4377             & 22.1H & 8.248 & 5.91  &  K2    & 4900  &  \nodata  & 0.30  & 0.31  & S & NC10,E11 \\
    \enddata
    \tablecomments{
$^{a}$ Distances marked with an H are from Hipparcos parallax measurements, those marked with an asterisk are assumed to be that of
their companion, and those marked with a c are calculated based on an assumed radius (see Section 2). \\
$^{b}$ Values in brackets indicate measurements that may be biased by a spatially unresolved companion. \\
$^{c}$ Spectral types followed by a colon are estimated from B-V color (see Section 2)\\
$^{e}$ Multiplicity abbreviations: S is Single, B is Binary, T is Triple, and Q is Quadruple\\
$^{f}$ Separation References: 2M = 2MASS; B07 = Biller et al. (2007); B10 = Biller et al. (2010); E11 = Evans et al. (2011); K07 = Kasper et al. (2007); L07 = Lafreniere et al. (2007); M90 = McAlister et al. (1990); NC10 = Nielson \& Close 2010; 
T08 = Torres et al. (2008); W11 = Wahhaj et al. (2011)
}
\end{deluxetable}


\begin{deluxetable}{rrrrrrrrrrrlr}
  \rotate
  \tabletypesize{\scriptsize}
  \tablewidth{0pt}
  \tablecaption{Proposed $\beta$ Pictoris Moving Group Members  \label{tab:betapic}}
  \tablehead{
    \multicolumn{3}{c}{} &
    \colhead{Dist$^{a}$} & 
    \colhead{$V^{b}$} & 
    \colhead{$K^{b}$} &
    \colhead{Prev.$^{c}$} & 
    \colhead{$T_\textrm{eff}$} & 
    \colhead{New} & 
    \colhead{Lum$^{b}$} & 
    \colhead{$\theta^{b}$} & 
    \colhead{Known$^{e}$} & 
    \colhead{Mult$^{f}$} \\
    \colhead{HD} & 
    \colhead{HIP}  &
    \colhead{Other} &
    \colhead{(pc)} & 
    \colhead{mag} &
    \colhead{mag} & 
    \colhead{SpT} &               
    \colhead{(K)} & 
    \colhead{SpT,$T_\textrm{eff}$} &  
    \colhead{($L_{\odot}$)} & 
    \colhead{(mas)} & 
    \colhead{Mult} & 
    \colhead{Ref}  }  \startdata
    203   & 560   &  HR9                        & 39.1H & 6.173 & 5.24  &    F3  & 6730  &    \nodata  & 4.5  & 0.37  & S & E11 \\
        \nodata  &  \nodata  & BD+17 232 A              & 52.6  & 10.584 & [6.72] &  K3    & 4730  &  \nodata  & 0.20  & 0.12  & B     & \nodata \\
        \nodata  &  \nodata  & BD+17 232 B          & 52.6* & 10.643 & \nodata & K5:   & 4420  &  \nodata  & 0.23  & 0.14  & B 1\farcs8 & M00  \\
        \nodata  &  \nodata  & TYC 1186-706-1            & 60    & 10.841 & 7.34  &  K7.5   & 3940  &  K7,4000  & 0.35  & 0.19  & \nodata & \nodata \\
        \nodata  &  \nodata  & 1RXS J010711.7-193529 & 54    & 11.290 & 7.25  &  M1   & 3650  &  \nodata  & 0.27  & 0.23  & \nodata & \nodata \\
        14082 A & 10680 & BD+28 382 A                & 39.4H & 7.031 & 5.79  &  F5    & 6440  &  F8,6120  & 2.0  & 0.30  & B     & E11 \\
              B & 10679 & BD+28 382 B                & 34.0H & 7.755 & 6.26  &  G2    & 5860  &  G1,5970  & 0.81  & 0.23  & B 13\farcs8 & 2M \\
       \nodata   & 11152 & LP 353-51                   & 30.7H & 11.307 & 7.35  &  M3    & 3350  &  \nodata  & 0.16  & 0.36  & \nodata & \nodata \\
    15115 & 11360 & BD+05 338                   & 44.8H & 6.781 & 5.82  &  F4   & 6580  &  F8,6190  & 3.42  & 0.33  & \nodata & \nodata \\
        \nodata  & 11437 A & AG Tri A                    & 42.3H & 10.1  & 7.08  &  K6   & 4260  &  \nodata  & 0.29  & 0.22  & B     & E11 \\
        \nodata  & B     & AG Tri B                   & 42.3* & 12.4  & 7.92  &  M2   & 3500  &  \nodata  & 0.070  & 0.16  & B 22\farcs1 & 2M,E11 \\
        \nodata  &  \nodata  & TYC 7558-655-1            & 35.7  & 10.417 & 7.23  &  K5   & 4395  &  \nodata  & 0.13  & 0.16  & \nodata & \nodata \\
        \nodata  & 12545 & BD+05 378          & 40.5H & 10.194 & 7.07  &  K6   & 4260  &  \nodata  & 0.25  & 0.21  & S     & B07,K07,NC10,E11 \\
    29391 A & 21547 & 51 Eri A & 29.8H & 5.209 & 4.54  &  F0    & 7200  &  \nodata  & 6.7  & 0.52  & T     & E11 \\
    Ba    &  Ba  & GJ 3305 A & 30    & [10.6] & [6.41] &  M1  & 3650  &  \nodata  & [0.16]  & [0.31] & T 66\farcs5 & 2M \\
    Bb    &  Bb  & GJ 3305 B & 30    & \nodata & \nodata &  \nodata  &  \nodata  &  \nodata  &  \nodata  &  \nodata  & T 0\farcs09 & K07 \\
        \nodata  & 23200 & V1005 Ori & 26.7H & 10.097 & 6.26  &  M0   & 3800  &  \nodata  & 0.17  & 0.33  & \nodata & \nodata \\
        \nodata  & 23309 & CD-57 1054 & 26.3H & 10.105 & 6.24  &  M0   & 3800  &  \nodata  & 0.16  & 0.33  & S     & B07,NC10 \\
        \nodata  & 23418 Aa & LP 476-207 Aa & 32.1H & [11.741] & [6.37] &  M3   & 3350  &  \nodata  & [0.11]  & [0.29] & B     & \nodata \\
        \nodata  & 23418 Ab & LP 476-207 Ab & 32.1* & \nodata & \nodata &  \nodata  &  \nodata  &  \nodata  &  \nodata  &  \nodata& B 1\farcs0 & T08 \\
        \nodata  &  \nodata  & BD-21 1074 A & 18    & 10.438 & 6.11  &  M1   & 3650  &  \nodata  & 0.056  & 0.31  & T     & \nodata \\
        \nodata  &  \nodata  & B         & 18*   & [11.500] & [6.12] &  \nodata  &  \nodata  &  \nodata  &  \nodata  &  \nodata  & T 8\farcs2 & 2M \\
        \nodata  &  \nodata  & C     & 18*   & \nodata & \nodata &  M3   & 3350  &  \nodata  &  \nodata  &  \nodata  & T 1\farcs2 & T08 \\
    35850 A & 25486 A & AF Lep A               & 26.8H & [6.296] & [4.93] &  F7    & 6270  &  \nodata  & [1.9]  & [0.40] & B     & E11 \\
    B     & B     & B                        & 26.8* & \nodata & \nodata & \nodata &  \nodata  &  \nodata  &  \nodata  &  \nodata  & B SB & T08 \\
        \nodata  &  \nodata  & V1311 Ori                & 36    & 11.098 & 7.01  &  M2   & 3500  &  \nodata  & 0.18  & 0.30  & \nodata & \nodata \\
    39060 & 27321 & $\beta$ Pic                & 19.3H & 3.848 & 3.53  &  A3    & 8720  &  \nodata  & 10 & 0.69  & \nodata & \nodata \\
    45081 & 29964 & AO Men                     & 38.5H & 9.989 & 6.81  &  K4   & 4590  &  \nodata  & 0.20  & 0.17  & S     & B07,K07,NC10 \\
    139084 A & 76629 A & V343 Nor A                  & 39.8H & 8.153 & 5.85  &  K0    & 5350  &  \nodata  & 0.89  & 0.26  & B     & NC10 \\
              B & B     & B                        & 39.8* & 14.8  & 9.19  &  M5   & 3000  &  \nodata  & 0.036  & 0.16  & B 10\farcs4 & 2M \\
        \nodata  &  \nodata  & J16430128-1754274 & 57    & \nodata & 8.55  &  M0.5  & 3700  &  \nodata  &  \nodata  &  \nodata  & \nodata & \nodata \\
       155555 A & 84586 A &  V824 Ara A     & 31.4H & [6.869] & [4.70] &  G7   & 5620  &  \nodata  & [1.6]  & [0.40] & T  & B07,NC10 \\
              B & B     &  V824 Ara B   & 31.4* & \nodata & \nodata &  \nodata  &  \nodata  &  \nodata  &  \nodata  &  \nodata  & T SB & B07,T08,NC10 \\
              C & C     &  V824 Ara C  & 31.4* & [12.8] & [7.63] &  M3   & 3350  &  \nodata  & [0.040]  & [0.18] & T 34\farcs0 & 2M,NC10 \\
        \nodata  &  \nodata  & GSC 8350-1924 A     & 76    & [13.47] & [7.99] &  M3   & 3350  &  \nodata  & [0.13]  & [0.13] & B     & \nodata \\
        \nodata  &  \nodata  & GSC 8350-1924 B & 76* & \nodata & \nodata &  \nodata  &  \nodata  &  \nodata  &  \nodata & \nodata  & B 0\farcs8 & T08 \\
        \nodata  &  \nodata  & CD-54 7336                & 66    & 9.529 & 7.36  &  K1    & 5080  &  \nodata  & 0.72  & 0.16  & \nodata & \nodata \\
    160305 & 86598 & CD-50 11467                & 76H   & 8.357 & 6.99  &  F9  & 6110  &  \nodata  & 2.3  & 0.17  & \nodata & \nodata \\
    161460 A &  \nodata  & CD-15 7414 A & 74    & [8.986] & [6.78] &  K0   & 5250  &  \nodata  & [1.4]  & [0.18] & B     & \nodata \\
    B     &  \nodata  & CD-15 7414 B & 74*   & \nodata & \nodata &  \nodata  &  \nodata  &  \nodata  &  \nodata  &  \nodata  & B 0\farcs1 & T08 \\
       164249 A & 88399 A & CD-51 11312 A              & 46.9H & 7.007 & 5.91  &  F6    & 6350  &  \nodata  & 3.0  & 0.28  & B     & \nodata \\
              B & 88399 B & CD-51 11312 B              & 46.9* & 12.5  & 8.27  &  M2   & 3500  &  \nodata  & 0.083  & 0.16  & B 6\farcs4 & 2M \\
    165189 A &  \nodata  & HR6749 A          & 44    & 5.624 & [6.39] &  A5    & 8200  &  \nodata  & 11  & 0.34  & B     & \nodata \\
    B     &  \nodata  & B                        & 44*   & 5.650 & \nodata & A8:   & 7550  &  \nodata  & 10  & 0.39  & B 1\farcs7 & T08 \\
    319139 A &  \nodata  & V4046 Sgr A & 73    & [10.681] & [7.25] &  K6   & 4260  &  \nodata  & [0.51]  & [0.17] & B     & \nodata \\
    B     &  \nodata  & V4046 Sgr B & 73*   & \nodata & \nodata & \nodata &  \nodata  &  \nodata  &  \nodata  &  \nodata  & B SB & T08 \\
        \nodata &  \nodata  & GSC 7396-0759             & 73    & 12.8  & 8.54  &  M1   & 3650  &  \nodata  & 0.12  & 0.11  & \nodata & \nodata \\
    168210 & 89829 & CD-29 14813                & 75.5H & 8.797 & 7.05  &  G5    & 5750  &  \nodata  & 1.6  & 0.16  & \nodata & \nodata \\
    172555 A & 92024 A & HR 7012                    & 29.2H & 4.767 & 4.30  &  A6  & 8350  &  \nodata  & 10  & 0.49  & T     & B07,NC10 \\
    B     & B     & CD-64 1208 A & 29*   & [9.5] & [6.10] &  K5   & 4400  &  \nodata  & [0.20]   & [0.25] & T 74\farcs0 & 2M,B07,NC10 \\
    C     & C     & CD-64 1208 B & 29*   & \nodata & \nodata &  \nodata  &  \nodata  &  \nodata  &  \nodata  &  \nodata  & T 0\farcs2 & T08 \\
        \nodata  &  \nodata  & TYC 9073-0762-1          & 54    & 12.216 & 7.85  &  M1   & 3650  &  \nodata  & 0.12  & 0.15  & \nodata & \nodata \\
        \nodata  &  \nodata  & CD-31 16041              & 51    & 11.301 & 7.46  &  K7   & 4000  &  \nodata  & 0.17  & 0.15  & \nodata & \nodata \\
    174429 A & 92680 A & PZ Tel A                   & 49.7H & 8.406 & 6.37  &  G9   & 5410  &  \nodata  & 1.1  & 0.22  & B     & NC10 \\
    B     & B     & PZ Tel B     & 49.7* & \nodata & \nodata &  \nodata  &  \nodata  &  \nodata  &  \nodata  &  \nodata  & B 0\farcs3 & B10,NC10 \\
        \nodata  &  \nodata  & TYC 6872-1011-1          & 79    & 11.897 & 8.02  &  M0   & 3800  &  \nodata  & 0.28  & 0.14  & \nodata & \nodata \\
        \nodata  &  \nodata  & CD-26 13904 A & 80    & [10.203] & [7.37] &  K4  & 4590  &  \nodata  & [0.67]  & [0.15] & B     & \nodata \\
        \nodata  &  \nodata  & CD-26 13904 B & 80*   & \nodata & \nodata & \nodata &  \nodata  &  \nodata  &  \nodata  &  \nodata  & B 1\farcs1 & T08 \\
    181296 A & 95261 A &  $\eta$ Tel A             & 47.7H & 5.015 & 5.01  &  A0    & 9520  &  \nodata  & 26 & 0.36  & B     & \nodata \\
              B & B     &  $\eta$ Tel B  & 47.7* & \nodata & \nodata &  \nodata  &  \nodata  &  \nodata  &  \nodata  &  \nodata  & B 4\farcs2 & T08 \\
    181327 & 95270 &  $\eta$ Tel C & 50.6H & 7.037 & 5.91  &  F6    & 6350  &  \nodata  & 3.4  & 0.28  & \nodata & \nodata \\
        \nodata  &  \nodata  &  TYC 7443-1102-1  & 58    & 11.797 & 7.85  &  M0.0   & 3800  &  K7.5,3900  & 0.15  & 0.14  & B & \nodata \\
        \nodata  &  \nodata  & 1RXS J195602.8-320720 & 58    & 13.3  & 8.11  &  M4   & 3150  &  M3.5,3250  & 0.11  & 0.17  & B 26\farcs3  & 2M \\
        \nodata  &  \nodata  & 1RXS J200136.9-331307 & 62    & 12.3  & 8.24  &  M1   & 3650  &  \nodata  & 0.14  & 0.14  & \nodata & \nodata \\
    191089 & 99273 &  CD-26 14819                & 53.5H & 7.181 & 6.08  &  F5    & 6440  &  \nodata  & 3.2  & 0.25  & \nodata & \nodata \\
    196982 A & 102141 A &  AT Mic A                  & 10.2H & 11.0  & [4.94] &  M4   & 3150  &  \nodata  & 0.036  & 0.60  & B     & B07,NC10 \\
    B     & B     &  AT Mic B                & 10.2* & 11.1  & \nodata &  M4  & 3150  &  \nodata  & 0.033  & 0.57  & B 3\farcs3 & B07,T08,NC10 \\
    197481 & 102409 &  AU Mic                    & 9.94H & 8.757 & 4.53  &  M1   & 3650  &  \nodata  & 0.10  & 0.72  & S     & B07,E11 \\
    199143 A & 103311 A & BD-17 6127 A & 47.7H & [7.318] & [5.81] & F7   & 6270  &  \nodata  & [2.4]  & [0.25] & B     & NC10 \\
    B     & B     & B          & 47.7* & \nodata & \nodata & \nodata & \nodata & \nodata &  \nodata  &  \nodata  & B 1\farcs1 & T08,NC10 \\
    358623 A &  \nodata & AZ Cap A                  & 47    & [10.625] & [7.04] &  K6   & 4260  &  \nodata  & [0.22]  & [0.17] & B     & E11 \\
    B     &  \nodata  & AZ Cap B           & 47*   & \nodata & \nodata &  \nodata  &  \nodata  &  \nodata  &  \nodata  &  \nodata  & B 2\farcs2 & T08 \\
        \nodata  &  \nodata  & TYC 2211-1309-1          & 46    & 11.367 & 7.72  &  M0   & 3800  &  K7,4000  & 0.13  & 0.15  & \nodata & \nodata \\
        \nodata  &  \nodata  & CP-72 2713               & 36    & 10.533 & 6.89  &  K7   & 4000  &  \nodata  & 0.17  & 0.22  & \nodata & \nodata \\
        \nodata  & 112312 A &  WW PsA                    & 23.6H & 12.1  & 6.93  &  M4   & 3150  &  \nodata  & 0.075  & 0.36  & B     & B07,NC10,E11 \\
        \nodata  & B     &  TX PsA                   & 20    & 13.4  & 7.79  &  M5   & 3000  &  \nodata  & 0.032  & 0.31  & B 35\farcs9 & 2M,E11 \\
        \nodata  &  \nodata  &  BD-13 6424               & 28    & 10.685 & 6.57  &  M0   & 3800  &  \nodata  & 0.11  & 0.25  & \nodata & \nodata \\
 
    \enddata
    \tablecomments{
$^{a}$ Distances marked with an H are from Hipparcos parallax measurements, those marked with an asterisk are assumed to be that of
their companion, and those marked with a c are calculated based on an assumed radius (see Section 2). \\
$^{b}$ Values in brackets indicate measurements that may be biased by a spatially unresolved companion. \\
$^{c}$ Spectral types followed by a colon are estimated from B-V color (see Section 2)\\
$^{e}$ Multiplicity abbreviations: S = Single, B = Binary, T = Triple, Q = Quadruple \\
$^{f}$ Separation References: 2M = 2MASS; B07 = Biller et al. (2007); B10 = Biller et al. (2010); E11 = Evans et al. (2011) K07 = Kasper et al. (2007); 
L07 = Lafreniere et al. (2007); M00 = Morlet et al. (2000); NC10 = Nielson \& Close 2010;  T08 = Torres et al. (2008)
}
\end{deluxetable}

\begin{deluxetable}{rrrrr}
  \tabletypesize{\scriptsize}
  \tablewidth{0pt}
  \tablecaption{Li[I] and H$\alpha$ Equivalent widths \label{tab:LiandHa}}
  \tablehead{
    \colhead{Star} & 
    \colhead{ } & 
    \colhead{EW[Li\,I]}  & 
    \colhead{EW[H$\alpha$]} & 
    \colhead{ } \\ 
    \colhead{Name} &
    \colhead{SNR} &
    \colhead{\AA} &
    \colhead{\AA} &
    \colhead{MG} }  
    \startdata
      TYC 1752-63-1$^{8}$ & 100   & 0.09  & -1.41 & AB \\
    IS Eri $^{8}$ & 240   & 0.17  & 1.14  & AB \\
    HIP 14809$^{8}$ & 275   & 0.14  & 0.94  & AB \\
    HD 25457$^{8}$ & 200   & 0.09  & 1.18  & AB \\
    HD 25953$^{8}$ & 175   & 0.11  & 1.41  & AB \\
    HIP 51317$^{5}$ & 60    & $<$0.04  & 0.38  & AB \\
    PX Vir$^{5}$ & 240   & 0.13  & 0.85  & AB \\
    HD 152555$^{5,8}$ & 230   & 0.12  & 1.15  & AB \\
    HD 160934$^{8}$ & 125   & $<$0.04 & -0.95 & AB \\
    HD 199058$^{8}$ & 150   & 0.16  & 1.08  & AB \\
    LO Peg$^{8}$ & 350   & 0.15  & -0.6  & AB \\
    HIP 114066$^{8}$ & 100   & $<$0.04 & -2    & AB \\    
    TYC 1186-706-1$^{8}$ & 130   & 0.37  & -0.65 & $\beta$ \\
    HIP 10679$^{8}$ & 215   & 0.16  & 0.9   & $\beta$ \\
    HIP 10680$^{8}$ & 175   & 0.11  & 1.59  & $\beta$ \\
    HD 15115$^{8}$ & 175   & 0.1   & 0.74  & $\beta$ \\
    TYC 7443-1102-1$^{8}$ & 100   & 0.11  & -0.88 & $\beta$ \\
    1RXS J19506.8-3320720$^{8}$ & 40    &  $<$0.5     & -5    & $\beta$ \\
    TYC 2211-1309-1$^{8}$ & 150   & $<$0.04 & -1.72 & $\beta$ \\
\enddata
 \tablecomments{ $^{5}$ Observed during the May run.\\
 $^{8}$ Observed during the August run.}
\end{deluxetable}

\clearpage


\begin{thebibliography}{}

\bibitem[Bailey et al. (2011)]{bai2011} Bailey, J. I., III, White, R. J., Blake, C. H., 
	Charbonneau, D., Barman, T., Tanner, A. M. \& Torres, G. 2011, \apj, submitted
\bibitem[Baines et al.(2007)]{bai2007} Baines, E. K., van Belle, G. T., 
	ten Brummelaar, T. A., McAlister, H. A., Swain, M., Turner, N. H., 
	Sturmann, L., \& Sturmann, J. 2007, \apj, 661, 195
\bibitem[Baines et al. (2008)]{bai2008} Baines, E. K., McAlister,H. A., ten 
	Brummelaar, T. A., Turner,N. H., Sturmann, J., Sturmann, L., Goldfinger, 
	P. J., \& Ridgway, S. T. 2008a, \apj, 680, 728
\bibitem[Baines et al.(2010)]{bai2010} Baines, E. K., et al. 2010, \apj, 710, 1365
\bibitem[Baraffe et al.(1998)]{bar1998} Baraffe, I., Chabrier, G., Allard, F., \& 
	Hauschildt, P. H. 1998, A\&A, 337, 403
\bibitem[Baraffe et al.(2002)]{bar2002} Baraffe, I., Chabrier, G., Allard, F., \& 
	Hauschildt, P. H. 2002, A\&A, 382, 563
\bibitem[Bessell (2000)]{bes2000} Bessell, M. S. 2000, PASP, 112, 773
\bibitem[Biller et al.(2007)]{bil2007} Biller, B. A., et al. 2007, \apj, 173, 143
\bibitem[Biller et al.(2010)]{bil2010} Biller, B. A., et al. 2010, \apj, 720, 82
\bibitem[Bodenheimer (1965)]{bod1965} Bodenheimer, P. 1965, \apj, 142, 451
\bibitem[Boden (2007)]{bod2007} Boden, A. F. 2007, New Astronomy Reviews, 51, 617
\bibitem[Boyajian et al.(2008)]{boy2008} Boyajian, T. S., et al. 2008, \apj, 683, 424
\bibitem[Colavita et al.(1999)]{col1999} Colavita, M. M., et al. 1999, \apj, 510, 505
\bibitem[Collier-Cameron \& Foing (1997)]{cam1997} Collier-Cameron, C. A., \& Foing, B. H. 1997, Observatory, 117, 218
\bibitem[Coud\'{e} du Foresto et al.(1997)]{cou1997}Coud\'{e} du Foresto, V., Ridgway, S., \& Mariotti, J. M. 1997, A\&A, 121, 379 
\bibitem[Close et al.(2005)]{clo2005} Close, L., M., et al. 2005, Nature, 433, 286
\bibitem[Covino et al.(1997)]{cov1997} Covino, E., Alcala, J. M., Allain, S.,
			 Bouvier, J., Terranegra, L., \& Krautter, J. 1997, A\&A, 328, 187
\bibitem[Craig et al.(1997)]{cra1997} Craig, N., Christian, D. J., Dupuis, J., \& Roberts, B. A. 1997, \aj, 114, 244
\bibitem[da Silva et al.(2009)]{sil2009} da Silva, L., Torres, C. A. O., de La Reza, R.,
			Quast, G. R., Melo, C. H. F., Sterzik, M. F. 2009, A\&A, 508, 833
\bibitem[Davis et al.(1999)]{dav1999} Davis, J., Tango, W. J., Booth, A. J., 
			ten Brummelaar, T. A., Minard, R. A., \& Owens, S. M. 1999, \mnras, 303, 703
\bibitem[Evans et al. (2011)]{eva2011} Evans, T. M., et al. 2012, \apj, accepted
\bibitem[Feigelson et al.(2006)]{fei2006} Feigelson, E. D., 
			Lawson, W. A., Stark, M., Townsley, L., \& Garmire, G. P. 2006,
			\aj, 131, 1730
\bibitem[Glushneva et al.(2002)]{glu2002} Glushneva, I. N., Shenavrin, V. I., \& Roshchina, I. A. 2002,
			Astronomical and Astrophysical Transactions, 21, 317
\bibitem[Gray et al.(2001)]{gra2001} Gray, R. O., Napier, M. G., \& Winkler, L. I. 2001,
			\aj, 121, 2148
\bibitem[Gray et al.(2003)]{gra2003} 	Gray, R. O., Corbally, C. J., Garrison, R. F., McFadden, M. T., \& Robinson, P. E. 2003,
			\aj, 126, 2048
\bibitem[Hanbury Brown et al.(1974)]{han1974} Hanbury Brown, R., Davis, J., Lake, R. J. W.,
			\& Thompson, R. J. 1974, \mnras, 167, 475
\bibitem[Haniff (2007)]{han2007} Haniff, C. 2007, New Astronomy Reviews, 51, 583
\bibitem[Hartigan et al.(1994)]{har1994} Hartigan, P., Strom, K. M., \& Strom, S. E. 1994,
			\apj, 427, 961
\bibitem[Hillenbrand \& White (2004)]{white2004} Hillenbrand, L. A., \& White, R. J. 2004,
		 	\apj, 604, 741
\bibitem[H$\o$g et al.(2000)]{hog2000} H$\o$g, E., et al. 2000, A\&A, 355, 27
\bibitem[Hormuth et al.(2007)]{hor2007} Hormuth, F., Brandner, W., Hippler, S., Janson, M., \& Henning, T.
			2007, A\&A, 463, 707
\bibitem[Hummel et al.(2003)]{hum2003} Hummel, C. A., et al. 2003, \aj, 125, 2630
\bibitem[Janson et al. (2007)]{jan2007} Janson, M., et al. 2007, \aap, 462, 615
\bibitem[Kasper et al.(2007)]{kas2007} Kasper, M.,  Apai, D., Janson, M., \& Brandner, W.
			2007, A\&A, 472, 321
\bibitem[Kenyon \& Hartmann (1995)]{ken1995} Kenyon, S. J., \& Hartmann, L. 1995, \apjs, 101, 117
\bibitem[Kirkpatrick et al.(1991)]{kir1991} Kirkpatrick, J. D., Henry, T. J., \& McCarthy, D. W.
			1991, \apjs, 77, 417
\bibitem[Kiss et al.(2011)]{kis2011} Kiss, L. L., et al. 2011, \mnras, 411, 117
\bibitem[Labeyrie (1975)]{lab1975} Labeyrie, A. 1975, \apj, 196, 71
\bibitem[L\'{e}pine \& Simon (2009)]{lep2009} L\'{e}pine, S., \& Simon, M. 2009,
			\aj, 137, 3632
\bibitem[Lobel (2007)]{lob2007} Lobel, A. 2007, SpectroWeb: An Interactive Graphical Database of 
	Digital Stellar Spectral Atlases, in "The Ultraviolet Universe: Stars from Birth to Death", 
	26th meeting of the IAU, Editorial Complutense Univ. of Madrid, ed. A. Gomez de Castro and 
	M. Barstow, 167, Aug 2006, Prague, Czech Republic. 
\bibitem[L\'{o}pez-Santiago et al.(2006)]{lop2006} L\'{o}pez-Santiago, J., Montes, D.,
			Crespo-Chacn, I., \& Fernndez-Figueroa, M. J. 2006, \apj, 643, 1160
\bibitem[Lowrance et al.(2000)]{low2000} Lowrance, P., et al. 2000, \apj, 541, 390
\bibitem[Luhman et al.(2003)]{luh2003} Luhman, K. L., Stauffer, J. R., Muench, A. A., Rieke, G. H., 
	Lada, E. A., Bouvier, J., \& Lada, C. J. 2003, \apj, 593, 1093
\bibitem[Luhman et al.(2005)]{luh2005} 	Luhman, K. L., Stauffer, J. R., \& Mamajek, E. E. 2005, \apj, 628, 69
\bibitem[Mamajek \& Hillenbrand (2008)]{mam2008} Mamajek, E. E., \& Hillenbrand, L. A. 2008, \apj, 687, 1264
\bibitem[Mathieu et al.(2007)]{mat2007} Mathieu, R. D., Baraffe, I., Simon, M., Stassun, K. G., \& White, R. 2007
			University of Arizona Press, Tucson, 951, 411
\bibitem[MacDonald \& Mullen (2010)]{mac2010} Macdonald, J., \& Mullan, D. J. 2010, \apj, 723, 1599
\bibitem[McAlister et al.(1990)]{mca1990} McAlister, H., Hartkopf, W. I., \& Franz, O. G. 1990, \aj, 99, 965
\bibitem[Mentuch et al.(2008)]{men2008} Mentuch, E., Brandeker, A.,	van Kerkwijk, M. H., Jayawardhana, R., 
	\& Hauschildt, P. H.	2008, \apj, 689, 1127
\bibitem[Michelson \& Pease (1921)]{mic1921} Michelson, A. A., \& Pease, F. G. 1921, \apj, 53, 249
\bibitem[Montes et al.(2001)]{mon2001} Montes, D., López-Santiago, J., G\'{a}lvez, M. C., Fern\'{a}ndez-Figueroa, 
	M. J., De Castro, E. \& Cornide, M. 2001, \mnras, 328, 48
\bibitem[Mo\'{o}r et al.(2006)]{moo2006} Moor, A., Abraham, P., Derekas, A., Kiss, C., 
			Kiss, L. L., Apai, D., Grady, C., \& Henning, T. 2006, \apj, 644, 525
\bibitem[Morlet \& Gili (2000)]{mor2000} Morlet, G., Salaman, M., \& Gili, R. 2000, A\&A, 145, 67
\bibitem[Mourard et al.(2009)]{mou2010} Mourard, D., et al. 2010, A\&A, 508, 1073
\bibitem[Nielson \& Close (2010)]{nie2010} Nielson, E. L., \& Close, L. M. 2010, \apj, 717, 878
\bibitem[Ortega et al.(2007)]{ort2007}	Ortega, V. G., Jilinski, E., de La Reza, R., \& Bazzanella, B.
			2007, \mnras, 377, 441
\bibitem[Perryman et al.(2009)]{per2009} Perryman, M. A. C., et al. 2009, A\&A, 500, 501
\bibitem[Petrov et al.(2007)]{pet2007} Petrov, R. G., et al. 2007, 
			A\&A, 464, 1
\bibitem[Raghavan et al.(2010)]{rag2010} Raghavan, D., et al. 2010, \apj, 190, 1
\bibitem[Schlieder et al.(2010)]{sch2010} Schlieder, J. E., L\'{e}pine, S.,
			\& Simon, M. 2010, \aj, 140, 119
\bibitem[Siess et al.(2000)]{sie2000} Siess, L., Dufour, E., \& Forestini, M. 2000,
			A\&A, 258, 593
\bibitem[Skrutskie et al.(2006)]{skr2006} Skrutskie, M.F., et al. 2006, \aj, 131, 1163
\bibitem[Song et al.(2003)]{son2003} Song, I., Zuckerman, B., \& Bessell, M. S. 2003, \apj, 599, 342
\bibitem[Strassmeier \& Fekel et al.(1990)]{str1990} Strassmeier, K. G., \& Fekel, F. C. 1990, A\&A, 230, 389
\bibitem[ten Brummelaar et al.(2005)]{ten2005} ten Brummelaar, T.A., et al. 2005,
			\aj, 628, 453 
\bibitem[Torres et al.(2006)]{tor2006} Torres, C. A. O., Quast, G. R., da Silva, L., 
			de La Reza, R., Melo, C. H. F., \& Sterzik, M. 2006, A\&A, 460, 695
\bibitem[Torres et al.(2008)]{tor2008} Torres, C. A. O., Quast, G. R., Melo, C. H. F., \& Sterzik, M. F.
			2008, Handbook of Star Forming Regions, Volume II: The Southern Sky ASP Monograph Publications, 5, 757
\bibitem[Valenti \& Fischer (2005)]{val2005}	Valenti, J. A., \& Fischer, D. A. 2005,
			\apjs, 159, 141
\bibitem[van Belle \& von Braun et al.(2009)]{bel2009} van Belle, G. T., \& von Braun, K. 2009,
			\apj, 694, 1085
\bibitem[van Leeuwen (2007)]{lee2007} van Leeuwen, F. 2007, A\&A, 474, 653
\bibitem[Wahhaj et al.(2011)]{wah2011} Wahhaj, Z., et al. 2011, \apj, 729, 139
\bibitem[Webb et al.(1999)]{web1999} Webb, R. A., Zuckerman, B., Platais, I., Patience, J., White, R. J., Schwartz, M. J., 
  \& McCarthy, C. 1999, \apj, 512, 63
\bibitem[Weise et al.(2010)]{wei2010} Weise, P., Launhardt, R., Setiawan, J., \& Henning, T. 2010, A\&A, 517, 88
\bibitem[West et al.(2008)]{wes2008} West, A., et al. 2008, \aj, 135, 785
\bibitem[White et al.(1999)]{whi1999} White, R. J., Ghez, A. M., Reid, I. N., \& Schultz, G. 1999, \apj, 520, 811
\bibitem[Wizinowich (2006)]{wiz2006} Wizinowich, P. L. 2006, SPIE, 6034, 1
\bibitem[Wyant (2002)]{wya2002} Wyant, J. C. 2002, SPIE, 4927, 154
\bibitem[Yee \& Jensen(2010)]{yee2010} Yee, J. C., \& Jensen, E. L. N. 2010,
			\apj, 711, 303
\bibitem[Young \& Arnett(2005)]{you2005} Young, P. A., \& Arnett, D. 2005, \apj, 618, 908
\bibitem[Zickgraf et al.(2005)]{zic2005} Zickgraf, F. J., Krautter, J., Reffert, S., Alcal\'{a}, J. M., Mujica, R., 
  Covino, E., \& Sterzik, M. F. 2005, A\&A, 433, 151
\bibitem[Zuckerman et al.(2001)]{zuc2001} Zuckerman, B., Song, I., Bessell, M. S., \& Webb, R. A.
			2001, \apj, 562, 87
\bibitem[Zuckerman \& Song (2004a)]{zuc2004a} Zuckerman, B., \& Song, I. 2004a,
			ARA\&A, 42, 685
\bibitem[Zuckerman et al.(2004b)]{zuc2004b}Zuckerman, B., Song, Inseok, \& Bessell, M. S. 2004b, \apj, 613, 65
\bibitem[Zuckerman et al.(2011)]{zuc2011} 	Zuckerman, B., Rhee, J. H., Song, I., \& Bessell, M. S. 2011, \apj, 732, 61	



\end{thebibliography}
\end{document}